\documentclass[a4paper,11pt]{article}

\usepackage{jcappub} 

\usepackage{bm}
\usepackage{url}
\usepackage{amsmath}
\usepackage{amsfonts}
\usepackage{morefloats}

\newcommand{\SolarProp}{{\sc HelioProp}}

\newcommand{\GeV}{{\rm GeV}}

\newcommand{\km}{{\rm km}}

\newcommand{\s}{{\rm s}}

\newcommand{\nT}{{\rm nT}}

\newcommand{\pbar}{{\bar p}}

\newcommand{\sigmav}{\langle \sigma_{\rm ann} v \rangle}

\title{Constraints on particle dark matter from cosmic-ray antiprotons}

\author[a,b]{N. Fornengo}
\author[c,d]{L. Maccione}
\author[a,b,e,f]{A. Vittino}

\affiliation[a]{Department of Physics, University of Torino, via P. Giuria 1, 10125 Torino, Italy}
\affiliation[b]{Istituto Nazionale di Fisica Nucleare, via P. Giuria 1, 10125 Torino, Italy}
\affiliation[c]{Ludwig-Maximilians-Universit\"{a}t, Theresienstra{\ss}e 37, D-80333 M\"{u}nchen, Germany}
\affiliation[d]{Max-Planck-Institut f\"{u}r Physik (Werner Heisenberg Institut), F\"{o}hringer Ring 6, D-80805 M\"{u}nchen, Germany}
\affiliation[e]{Institut de Physique Th\'eorique, CNRS URA 2306 \& CEA-Saclay, 91191 Gif-sur-Yvette,
France}
\affiliation[f]{Universit\'e Paris-Diderot, rue Thomas Mann, 75205 Paris cedex 13, France}

\emailAdd{fornengo@to.infn.it}
\emailAdd{luca.maccione@lmu.de}
\emailAdd{vittino@to.infn.it}

\abstract{Cosmic-ray antiprotons represent an important channel for dark matter indirect-detection
studies. Current measurements of the antiproton flux at the top of the atmosphere and theoretical
determinations of the secondary antiproton production in the Galaxy are in good agreement, with
no manifest deviation which could point to an exotic contribution in this channel. Therefore, antiprotons
can be used as a powerful tool for constraining particle dark matter properties. By using the spectrum of PAMELA data from 50 MV to 180 GV in rigidity, we derive bounds on the dark matter annihilation cross section (or decay rate, for decaying dark matter) for the whole spectrum of
dark matter annihilation (decay) channels and under different hypotheses of cosmic-rays
transport in the Galaxy and in the heliosphere. For typical models of galactic propagation, the
constraints are strong, setting a lower bound on the dark matter mass of a "thermal" relic
at about  40 -- 80 GeV for hadronic annihilation channels. These bounds are enhanced to about 150 GeV
on the dark matter mass, when large cosmic-rays confinement volumes in the Galaxy are considered,
and are reduced to  3-4 GeV for annihilation to light quarks (no bound for heavy-quark production) when
the confinement volume is small.
Bounds for dark matter lighter than few tens of GeV are due to the low energy part of the
PAMELA spectrum, an energy region where solar modulation is relevant: to this aim, we have
implemented a detailed solution of the transport equation in the heliosphere, which allowed us
not only to extend bounds to light dark matter, but also to determine the uncertainty on the constraints arising
from solar modulation modeling. Finally, we estimate the impact of soon-to-come
AMS-02 data on the antiproton constraints.
}

\begin{document}
\maketitle

\section{Introduction}

Several astronomical observations confirm the fact that the vast majority of the matter content of the Universe is in the form of an unknown component called dark matter (DM) \cite{Ade:2013zuv}. Among those DM candidates that are best motivated under a theoretical point of view, weakly interacting massive particles (WIMPs) play a special role: their weak interaction may allow them to possess
the correct relic abundance to explain the observed amount of dark matter and, at the same time,
lead to the possibility for WIMPs to produce observable astrophysical signals: gamma-rays, neutrinos,
electrons/positrons, antiprotons, antideuterons \cite{Donato:1999gy} and further indirect electromagnetic signals, in the whole electromagnetic spectrum down to radio frequencies.

Among the various channels for DM indirect detection, antiprotons are known to represent one of the
best options, since  the flux of cosmic antiprotons has been measured in recent years by many experimental collaborations to a good level of precision: BESS \cite{Orito:1999re,Maeno:2000qx}, AMS \cite{Aguilar:2002ad}, BESS-Polar \cite{Abe:2011nx} and PAMELA \cite{Adriani:2010rc,Adriani:2012paa}. Novel data are
expected from AMS-02.
On the theoretical side, antiprotons have been suggested for the first time as a possible signature of DM in \cite{Silk:1984zy,Stecker:1985jc} and then they have been studied as a way to constrain the properties of annihilating or decaying DM particles in a huge variety of theoretical frameworks starting from supersymmetry \cite{Jungman:1993yn,Bottino:1994xs,Bottino:1998tw,Bergstrom:1999jc,Donato:2003xg,Bottino:2005xy,Ferrer:2006hy,Bottino:2007qg,Cerdeno:2011tf,Belanger:2012ta,Ibarra:2008qg,Buchmuller:2009xv, Evoli:2011id,Delahaye:2013yqa} to Kaluza-Klein DM \cite{Bringmann:2005pp,Barrau:2005au,Hooper:2009fj} but also in relation to minimal DM models \cite {Cirelli:2008id} or, more recently, as a constraining signal for DM models with internal bremsstrahlung \cite{Garny:2011cj,Ibarra:2012dw,Chu:2012qy}.

 In this paper, our purpose is to derive updated constraints on the DM annihilation cross section (or lifetime in the case of decaying DM) from experimental measurements of the antiprotons flux at the top of the atmosphere in a completely model independent framework \cite{Donato:2008jk,Lavalle:2010yw,Kappl:2011jw,Garny:2012vt,Cirelli:2013hv}. In addition, and following the path traced in Ref.\cite{Fornengo:2013osa}, we wish to add to the analysis of antiproton
 bounds also a detailed modeling of solar modulation, which is a critical element for low antiproton
 energies, where most of the experimental data are available and which are the relevant energies
 to constrain light DM. In fact, for DM masses below 50 GeV the constraints come from antiprotons
with kinetic energies below 10 GeV, which is where solar modulation mostly affects the predicted fluxes.
Solid and meaningful constraints for light DM therefore require a detailed modeling of cosmic
rays transport in the heliosphere. We will therefore study in detail the way in which a charge dependent solar modulation can affect the  antiproton fluxes and the ensuing bounds. This will also allow us to quantify the impact of the uncertainties arising from solar modulation modeling.

 The novel information which can be gained by this analysis is therefore: i) determination of the
most updated bounds on DM properties from cosmic antiprotons, with the inclusion in the theoretical calculation of all the most relevant galactic transport phenomena, including reacceleration and
energy losses, and the use of the whole spectral information from the PAMELA data set, which allow
to set relevant bounds also on light dark matter particles; ii) determination of the impact of solar modulation modeling with explicit quantification
of the uncertainty arising from antiproton transport in the heliosphere.

The paper is organized as follows: Section \ref{sec:transport} very briefly summarizes the method used to describe the propagation of the antiprotons in our Galaxy. Section \ref{sec:solarmod} deals with the issue of solar modulation, by introducing the fully numerical method employed to model the transport of cosmic rays in the heliosphere. Section \ref{sec:Bounds} provides details about the way in which we calculate the bounds on the DM annihilation cross section (or decay rate). The bounds obtained from the PAMELA data are reported in Section \ref{sec:PAMELA}, while Section \ref{sec:AMS} shows
the projected sensitivity for future experiments, namely AMS-02. Section \ref{sec:Conclusions} summarizes our main conclusions.  

\section{Antiprotons production and propagation in the Galaxy}
\label{sec:transport}

\begin{table}[t]
\centering
\renewcommand{\arraystretch}{1.2}
  \begin{tabular}{ |l | c | c |}
\hline
    profile & $\rho(r,z)/\rho_\odot$ & parameters \\
\hline
    Isothermal  & $(1 + r_{\odot}^2/r_{s}^2 )/(1 + (r^2+z^2)/r_{s}^2)$ &  $r_s$ = 5 kpc\\  
    NFW          & $(r_{\odot}/\sqrt{r^2+z^2})(1 + r_{\odot}/r_{s})^2/ (1 + \sqrt{r^2+z^2}/r_{s})^2$  &  $r_s$ = 20 kpc\\  
    Einasto       & $\exp(-2[(\sqrt{r^2+z^2}/r_s)^{\alpha}-(r_{\odot}/r_s)^{\alpha}]/\alpha)$  &  $r_s$ = 20 kpc \; , \;$\alpha = 0.17$ \\  
\hline
    \end{tabular}
\caption{Dark matter density profiles $\rho(r,z)$ adopted in the present analysis.}
\label{tab:profiles}
\end{table}

Antiprotons can be produced in the Galaxy through two main mechanisms:
a primary flux is produced by DM in pair annihilation or decay events, while a secondary flux,
which represents the astrophysical background, is produced by the spallation of cosmic rays on the nuclei that populate the interstellar medium (ISM). 
  
Primary antiprotons are initially released in the ISM with an injected spectrum $dN_{\bar p}/dT$,
 where $T$ is the antiproton kinetic energy.  We model the spectrum by using the PYTHIA MonteCarlo event generator , for which we have adopted the version 8.160 \cite{Sjostrand:2007gs}.  After being produced, antiprotons propagate in the galactic environment
and are subject to a number of physics processes:  diffusion, energy losses, drifts and annihilations.
These processes can be described in terms of a transport equation, which we conveniently express here in cylindrical coordinates, i.e. a radial coordinate $r$ along the galactic disk and a vertical coordinate $z$ perpendicular to the disk:

\begin{equation}
\begin{split}
  -\nabla [K(r,z,E) \nabla n_{\bar{p}}(r,z,E)] + V_{c}(z)\frac{\partial}{\partial{z}}n_{\bar{p}}(r,z,E)+2h \delta (z)\Gamma^{\rm ann}_{\bar{p}}n_{\bar{p}}(r,z,E) +  \\
   \quad2h \delta (z) \partial_{E}(-K_{EE}(E)\partial_{E}n_{\bar{p}}(r,z,E)+b_{tot}(E)n_{\bar{p}}(r,z,E))= q_{\bar{p}}(r,z,E)
    \end{split}
\label{eq:transport}
  \end{equation}
 This equation governs the transport of both the primary component, produced by DM annihilation or decay (our signal) as well as the secondary component, due to cosmic rays interactions on the ISM (the background).
The first term of Eq. (\ref{eq:transport}) describes spatial diffusion, expressed through a diffusion coefficient $K(r,z,E)$  which we assume to be purely energy-dependent: 
\begin{equation}
K(r,z,E)=K(E)=\beta K_0 \cal{R} ^\delta
\end{equation} 
The
second term refers to convection away from the galactic plane  and $V_{c}$ denotes the convection velocity which we take to be constant and directed outwards along the z axis;
the third term describes the possibility that antiprotons annihilate on the gas present in the galactic
disk, with $\Gamma^{\rm ann}$ denoting the annihilation rate: 
\begin{equation}
\Gamma^{\rm ann}_{\bar{p}} = (n_{\rm H} + 4^{2/3}n_{\rm He})\sigma^{\bar{p}p}_{\rm ann}v_{\bar{p}}
\end{equation}
 The fourth term takes into account the diffusion mechanism in the momentum space known as reacceleration; this process is ruled by the momentum diffusion coefficient which is related to the spatial diffusion in this way: 
\begin{equation}
K_{EE}(E)~=~\frac{2}{9}V_{a}^2\frac{E^2\beta^2}{K(E)}
\end{equation} 
being $V_{a}$ the alfv\'{e}nic speed of the magnetic shock waves that are responsible for the reacceleration process. For this parameter, we assume a constant value. Lastly, the fifth term describes the energy loss mechanisms that antiprotons can undergo during their propagation such as ionization, Coulomb and adiabatic losses, as well as the energy drift due to reacceleration:
\begin{equation}
b_{tot}(E)~=~ \left.\left(\frac{dE}{dt}\right)\right|_{ion}~+~\left.\left(\frac{dE}{dt}\right)\right|_{Coul}~+~\left.\left(\frac{dE}{dt}\right)\right|_{adi}~+~b_{reac}(E)
\end{equation}    
being $b_{reac}(E)~=~\frac{(1+\beta^2)}{E} K_{EE}$ while the energy losses coefficients $\left.\left(\frac{dE}{dt}\right)\right|_i$ are the ones defined in \cite{Maurin:2002ua}.  
The source term appearing in the right-hand-side
is given by: 
\begin{equation}
   q_{\bar{p}}(r,z,E) = \frac{1}{2} \sigmav \frac {dN_{\bar{p}}}{dT}\left(\frac{\rho(r,z)}{m_{DM}}\right)^2
\label{eq:source1}
\end{equation}
for annihilating DM, and:
 \begin{equation}
  q_{\bar{p}}(r,z,E) =  \Gamma_{\rm dec} \frac {dN_{\bar{p}}}{dT}\frac{\rho(r,z)}{m_{DM}}    \label{eq:source2}
 \end{equation}
for decaying DM. In the previous equations, $\sigmav$ is the thermally averaged annihilation cross section, $\Gamma_{\rm dec}$  is the DM decay rate ($\Gamma_{\rm dec}=1/\tau$ with $\tau$ the DM lifetime), $\rho(r,z)$
is the DM density profile (in our analysis we will use the profiles listed in Table \ref{tab:profiles} and
we adopt a local DM density of 0.39 ${\mathrm{GeV}\: \mathrm{cm}^{-3}}$). 
To solve Eq. (\ref{eq:transport}) we use the fully analytical formalism of the two-zone diffusion model, which has been widely described in literature \cite{Maurin:2002ua,diffusion1,diffusion2,diffusion3,Donato:2001ms}. To briefly sum up, the solution can be found by assuming the diffusion to be confined inside a cylinder of radius $R = 20 \: \mathrm{kpc}$ and centered at the galactic plane with vertical half-thickness $L$. A thin disk coincident with the galactic plane and of vertical half-height $h = 100 \:\mathrm{pc}$ is the place where cosmic rays may interact with the ISM. In this framework, the solution to the transport equation can be found by expanding the antiproton number density in a Bessel series:  
\begin{equation}
n_{\bar{p}}^{(0)}(r,z,E) = \sum_{i} n_{i}^{(0)}(z,E)J_0\left(\frac{\zeta_i r}{R}\right)
\end{equation}
where $J_0$ is the zeroth-order Bessel function of the first kind and $\zeta_i$ are its zeros of index $i$, while the optional superscript (0), if present, will indicate the solution of Eq. \ref{eq:transport} without reacceleration and energy losses. As already stated in \cite{Fornengo:2013osa}, if we neglect reacceleration and energy losses,  the coefficients of this Bessel expansion can be written (at the Earth's position, {\it i.e.} at z=0) as: 
 \begin{equation}
n_i^0(E,z=0) = \frac{e^{-aL}\;y_i(L,E)}{B_i \sinh(S_iL/2)}
\end{equation} 
where we have defined: 
\begin{eqnarray}
&& a    = (V_c)/(2 K) \\
&& S_i = 2[a^2 + (\zeta_i/R)^2] \\
&& A_i = (V_c + 2h\Gamma^{\rm ann}_{\bar{p}})/(K S_i) \\
&& B_i = K \, S_i[A_i + \coth(S_i L/2)]
\end{eqnarray} 
and: 
\begin{equation}
y_i(z,E) = 2 \int_0^z dz' \, {\rm e}^{a(z-z')}\, \sinh\left[\frac{S_i(z-z')}{2}\right]q_i(z',E)
\end{equation}
being: 
 \begin{equation}
q_i(z,E) = \frac{2}{[J_1(\zeta_i)R]^2}\int_0^R dr\; r J_0\left(\frac{\zeta_i r}{R}\right) q_{\bar{p}}(r,z,E)
 \end{equation}
 To include in our solution also reacceleration and energy losses, one has to solve:
 \begin{equation}
n_i + \frac{2h}{B_i} \frac{d}{dE}\left[b_{tot}(E)n_i-K_{EE}(E)\frac{dn_i^{\bar{p}}}{dE}\right]~=~n_i^{0}
\label{eq:transport2}
\end{equation}
Following Appendix $\it B$ of Ref. \cite{Donato:2001ms}, this equation can be discretised and solved numerically in a grid of energy values $E_j$. Basically, once discretised, Eq. \ref{eq:transport2} has the form: 
\begin{equation}
{\cal A}
\left(
\begin{array}{c}
\vdots\\
n_i^{j-1}\\
n_i^j\\
n_i^{j+1}\\
\vdots\\
\end{array}
\right)
=
\left(
\begin{array}{c}
\vdots\\
n_i^{0,j-1}\\
n_i^{0,j}\\
n_i^{0,j+1}\\
\vdots\\
\end{array}
\right)
\end{equation} 
where the label $j$ indicates that the corresponding element has been evaluated at the energy $E_j$ and ${\cal A}$ denotes a matrix whose entries correspond to the discretised form of Eq. \ref{eq:transport2} (we address the reader to \cite{Donato:2001ms}, where their explicit form is reported). The coefficients $n_i^j$ can then be found by inverting ${\cal A}$ (task that can be done numerically) Once that the coefficients $n_i^j$ are found, the interstellar flux can be expressed in this way: 
\begin{equation}
\phi_{\bar{p}}(E_j) = \frac{\beta_{\bar{p}}}{4\pi}n_{\bar{p}}(r=r_{\odot},z=0,E_j)
\end{equation}. 
For the values of the astrophysical parameters that enter Eq. (\ref{eq:transport}), we adopt the three sets called MIN, MED and MAX, \cite{Donato:2003xg}, listed in Table \ref{tab:parameters}. 

For the secondary antiproton flux the source term takes into account the hadronic interactions of
primary cosmic rays on the ISM:
\begin{equation}
q_{\bar{p}}(r,z,T) = \sum _{j}^{ISM}\sum_{i}^{CRs} \int_{T_{\rm th}}^{\infty} dT_i\frac{d\sigma^{ij}}{dT}(T,T_i)\phi_{i}(T_i)
\end{equation}
where $\phi_{i}(T_i)$ the flux of the primary cosmic rays species $i$  impinging on the ISM nucleus $j$ with a kinetic energy $T_i$, while $T_{\rm th}$ represents the minimal kinetic energy necessary to the production of one antiproton. For the secondary background we rely to Ref. \cite{Donato:2008jk}. We will comment on this secondary component and its uncertainties in Section \ref{sec:Bounds}.

\begin{table}[t]
\centering
\renewcommand{\arraystretch}{1.2}
    \begin{tabular}{ |l | c | c | c | c | c|}
\hline
     & $\delta$ & $ K_0$ (kpc$^2$/Myr) & $L$ (kpc) &  $V_c$ (Km/s) &  $V_a$ (Km/s)\\ 
\hline
    MIN & 0.85  & 0.0016 & 1 & 13.5 & 22.4\\  
    MED & 0.70  & 0.0112 & 4 & 12 &  52.9\\
    MAX & 0.46  & 0.0765 & 15 & 5 &  117.6 \\
\hline
    \end{tabular}
\caption{Set of parameters of the galactic propagation models for charged cosmic rays employed in the analysis \cite{diffusion1,Donato:2003xg}.}
\label{tab:parameters}
\end{table}
\section{Antiproton propagation in the heliosphere: solar modulation}
\label{sec:solarmod}
 

Before they are detected at Earth, CRs lose energy due to the solar wind while diffusing in the solar system \cite{Gleeson_1968ApJ}. This modulation effect depends, via drifts in the large scale gradients of the solar magnetic field (SMF), on the particle's charge including its sign \cite{1996ApJ...464..507C}. Therefore, it depends on the polarity of the SMF, which changes periodically every $\sim$11 years \cite{wilcox}. Besides the 11 year reversals, the SMF has also opposite polarities in the northern and southern hemispheres: at the interface between opposite polarity regions, where the intensity of the SMF is null, a heliospheric current sheet (HCS) is formed (see e.g.~ Ref. \cite{1981JGR....86.8893B}). The HCS swings in a region whose angular extension is described phenomenologically by the tilt angle $\alpha$. The magnitude of $\alpha$ depends on solar activity. Since particles crossing the HCS suffer from additional drifts because of the different orientation of the magnetic field lines, the intensity of the modulation depends on the extension of the HCS. This picture explains, at least qualitatively, the annual variability and the approximate periodicity of the fluctuations of CR spectra below a few GeV. 

The propagation of CRs in the heliosphere can be described by the following transport equation \cite{1965P&SS...13....9P}:
\begin{equation}
\frac{\partial f}{\partial t} = -(\vec{V}_{\rm sw}+\vec{v}_d)\cdot \nabla f + \nabla\cdot (\bm{H}\cdot\nabla f) + \frac{P}{3}(\nabla\cdot\vec{V}_{\rm sw})\frac{\partial f}{\partial P}\;,
\label{eq:solartransport}
\end{equation}
where $f$ represents the CR phase space density, averaged over momentum directions,  $\bm{H}$ represents the (symmetrized) diffusion tensor,  $\vec{V}_{\rm sw}$ the velocity of the solar wind, $\vec{v}_{d}$ the divergence-free velocity associated to drifts, $P$ the CR momentum. The transport equation is solved in a generic 3D geometry within the heliosphere, with a  boundary  at 100~AU (see \cite{Bobik:2011ig} and Refs.~therein). The CR interstellar flux is given as a boundary condition and we assume that no sources are present within the solar system at the energies relevant to this work.

A model for solar propagation is therefore specified by fixing the solar system geometry, the properties of diffusion and those of winds and drifts. 
We describe the solar system diffusion tensor by  $\bm{H}= {\rm diag}(H_{\|}, H_{\perp r}, H_{\perp \theta})$, where the parallel $\|$ and  perpendicular $\perp$ {\rm components} are set with respect to the direction of the local magnetic field. We assume no diffusion in the perpendicular and azimuthal directions and we describe as drifts the effect of possible antisymmetric components in  $\bm{H}$. For the CR mean-free-path parallel to the magnetic field we take $\lambda_{\|} = \lambda_{0}(\rho/1~\GeV)(B/B_{\bigoplus})^{-1}$, 
 where $\rho$ denotes the rigidity, $B$ is the magnetic field and $B_{\bigoplus}$ is its normalization value at the Earth position, for which we adopt $B_{\bigoplus}=5~\nT$ according to \cite{2011ApJ...735...83S,2012Ap&SS.339..223S}.  
For $\rho < 0.1~\GeV$, $\lambda_{\|}$ does not depend on rigidity. 
We then compute $ H_{\|} = \lambda_{\|}v/3$. Perpendicular diffusion is assumed to be isotropic. According to numerical simulations, we assume $\lambda_{\perp r,\theta} = 0.02\lambda_{\|}$ \cite{1999ApJ...520..204G}. 

For the SMF, we assume a Parker spiral, although more complex geometries might be more appropriate for periods of intense activity:
\begin{equation}
\vec{B} = AB_{0}\left(\frac{r}{r_{0}}\right)^{-2}\left(\hat{r} - \frac{\Omega r\sin\vartheta}{V_{\rm sw}}\hat{\varphi}\right)\;,
\end{equation}
where $\Omega$ is the solar differential rotation rate, $\vartheta$ is the colatitude, $B_{0}$ is a normalization constant such that $B_{\bigoplus}=5~\nT$ and $A=\pm H(\theta-\theta')$ determines the magnetic field polarity through the $\pm$ sign. The presence of a HCS is taken into account by the Heaviside function $H(\theta-\theta')$. The HCS angular extent is described by the function $\vartheta' = \pi/2 + \sin^{-1}\left(\sin\alpha\sin(\varphi+\Omega r/V_{\rm sw})\right)$, where $0<\alpha<90^{\circ}$ is the tilt angle. The drift processes, due to magnetic irregularities and to the HCS, are related to the antisymmetric part  $H_{A}$ of the diffusion tensor as \cite{1977ApJ...213L..85J}:
\begin{equation}
\vec{v}_{\rm d} = \nabla\times( H_{A}\vec{B}/|B|) = {\rm sign}(q)v/3\vec{\nabla}\times\left(r_{L}\hat{B}\right)\;,
\label{eq:drifts}
\end{equation}
where $ H_{A} = pv/3qB$, $r_{L}$ is the particle's Larmor radius and $q$ is the charge. We refer to Refs. \cite{2011ApJ...735...83S,2012Ap&SS.339..223S} for more details on the implementation of the HCS and of drifts. Adiabatic energy losses due to the solar wind expanding radially at $V_{\rm sw}\sim400~\km/\s$ are taken into account.

 Eq.~(\ref{eq:solartransport}) expresses the fact that CRs lose energy adiabatically, due to the expansion of the solar wind, while propagating in the heliosphere. It is straightforward to notice that the larger their diffusion time (i.e.~the shorter their mean-free-path) the more energy they lose in propagation. This fact is at the basis of the simplest modulation model used in the literature, the so called force-field model \cite{Gleeson_1968ApJ}. In this picture, the heliospheric propagation is assumed to be spherically symmetric, and energy losses are described by the modulation potential $\Phi \propto |{\bm{H}}|/V_{\rm sw}$ and $\Phi$ is to be fitted against data. However, this model completely neglects the effects of $\vec{v}_{d}$, which may significantly alter the propagation path.
 Polarity $A$ and tilt-angle $\alpha$ are of particular importance in this respect. If $q\cdot A<0$, drifts force CRs to diffuse in the region close to the HCS, which enhances their effective propagation time and therefore energy losses, while if $q\cdot A>0$ drifts pull CRs outside the HCS, where they can diffuse faster \cite{2011ApJ...735...83S,2012Ap&SS.339..223S}. 
 As this is the only effect that depends on the charge-sign in this problem, and given that the force-field model does not account for it, the latter model cannot be properly used to describe CR spectra below a few GeV, where charge-sign effects  might be relevant \cite{1996ApJ...464..507C,GastSchael,Bobik:2011ig,DellaTorre:2012zz,
 Potgieter:2013cwj,Maccione:2012cu}.

In our analysis, we adopt the approach recently developed in the numerical program \SolarProp\ \cite{Maccione:2012cu} for the 4D propagation of CRs in the solar system.
We follow the stochastic differential equation approach described in Refs. \cite{2011ApJ...735...83S,2012Ap&SS.339..223S,2007JGRA..11208101A}. The cosmic-rays phase-space density is computed by sampling and averaging upon pseudo-particle trajectories, which are the result of a deterministic component related to the drifts, and of a random walk component, whose amplitude is sampled according to the local diffusion tensor \cite{gardiner2009stochastic,2012CoPhC.183..530K}.
Pseudo-particles injected at the Earth position are retro-projected in time inside the solar system until they reach the heliopause, where their properties are recorded. The local interstellar flux, which is effectively a boundary condition for this problem, is then used as an appropriate weight to determine the Earth spectrum. More details  on the actual numerical scheme are discussed in Refs. \cite{2011ApJ...735...83S,2012Ap&SS.339..223S,2007JGRA..11208101A}.

The solar modulation models adopted in our analyses are tuned for the PAMELA-data taking period
by using data on solar activity and on independent analyses on cosmic-rays derived in the
same propagation model \cite{Maccione:2012cu}. The tilt angle $\alpha$ for the PAMELA period (around a minimum of solar activity) has been determined to be around $20^\circ$ \cite{Maccione:2012cu,Potgieter2003,Potgieter2004,wilcox}, a value which we will adopt in our analysis. The polarity of the Sun magnetic field is negative \cite{Maccione:2012cu}. For the mean free path
$\lambda$ we adopt a few representative values (0.15 AU, 0.20 AU and 0.25 AU), which
are compatible with both the measured electron mean-free-paths and with proton mean-free-path inferred from neutron monitor counts and the solar spot number  \cite{Maccione:2012cu,Bobik:2011ig,Droge2005532}. For the spectral index $\gamma$, for definiteness we consider  $\gamma=1$, as  recently derived in the literature \cite{Potgieter:2013cwj,Loparco:2013pea}.

\begin{figure}[t]
\centering
\includegraphics[width=\textwidth]{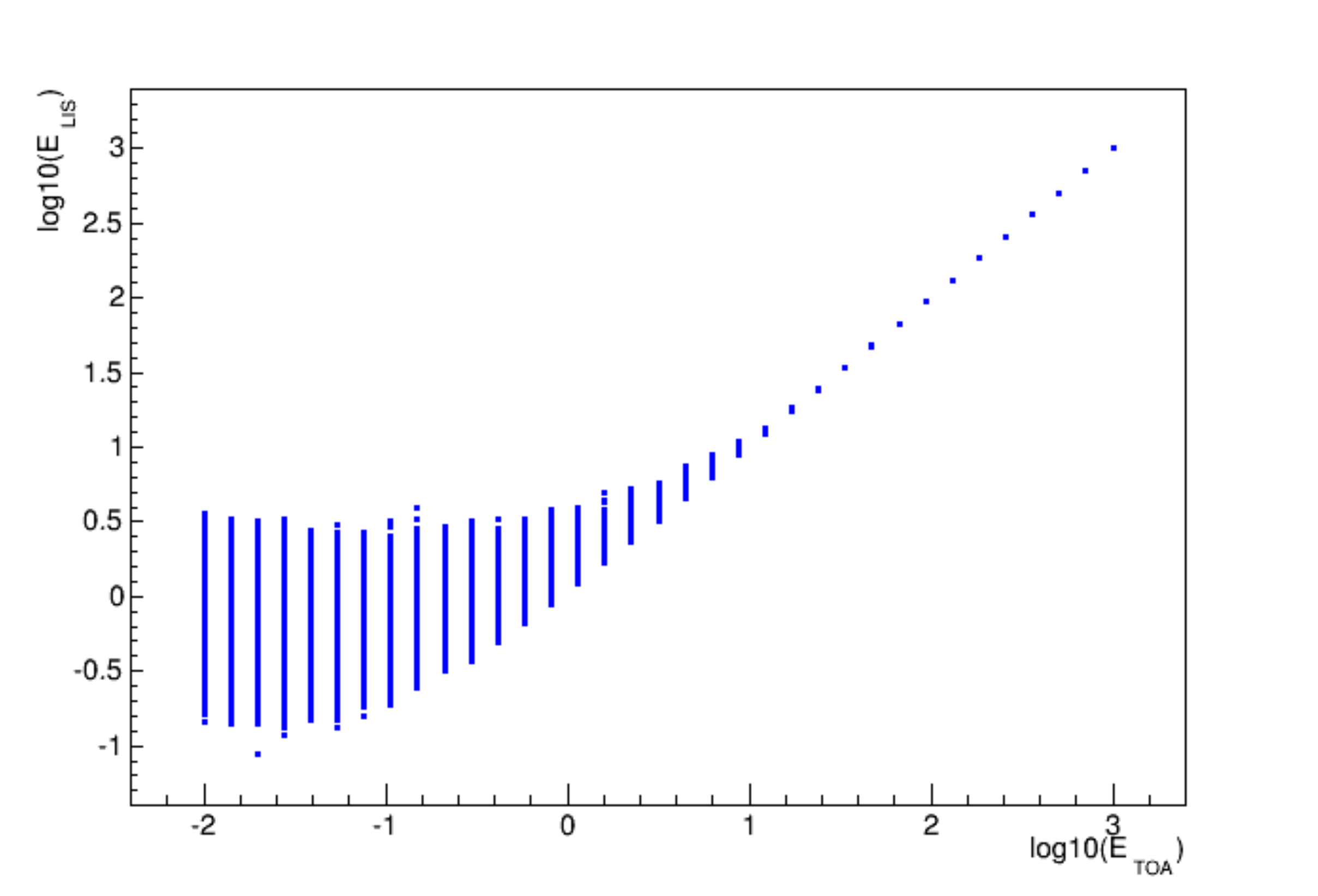}
\caption{Antiprotons local interstellar (LIS) kinetic energy per nucleon corresponding to a given top-of-atmosphere (TOA) kinetic energy per
nucleon, as obtained from a Monte Carlo modeling of cosmic-rays transport in the heliosphere.
The plot refers to
a solar-modulation model with a tilt angle $\alpha=20^{\circ}$ and a mean free path $\lambda=0.15~{\rm AU}$. Energy losses are negligible above about $50~{\rm GeV/n}$.}
\label{fig:sseloss}
\end{figure}

The main effects of solar system propagation on antiprotons are demonstrated in Fig.~\ref{fig:sseloss}, where we show how the TOA energy of these particles corresponds to the LIS energy of the same particle, for a sample of $10^{4}$ particles generated at each $E_{\rm TOA}$ in \SolarProp. While at high energy $E_{\rm LIS}=E_{\rm TOA}$, because diffusion is so fast that no energy losses occur, at low energies, below a few GeV, $E_{\rm LIS}>E_{\rm TOA}$ and the actual energy lost during propagation can vary significantly from particle to particle in our sample. This is due to the fact that energy losses are a function of the actual path, and the path is determined by a combination of drifts and random walks, being in fact a stochastic variable. Operationally, the flux observed at Earth at $E_{\rm TOA}$ is determined as a proper weighted average of the LIS flux at the energies $E_{\rm LIS}$ corresponding to that $E_{\rm TOA}$, as in Fig.~\ref{fig:sseloss}. 

\section{Antiproton fluxes and determination of the bounds on DM properties}
\label{sec:Bounds}

\begin{figure}[t]
\centering
\includegraphics[width=0.45\textwidth]{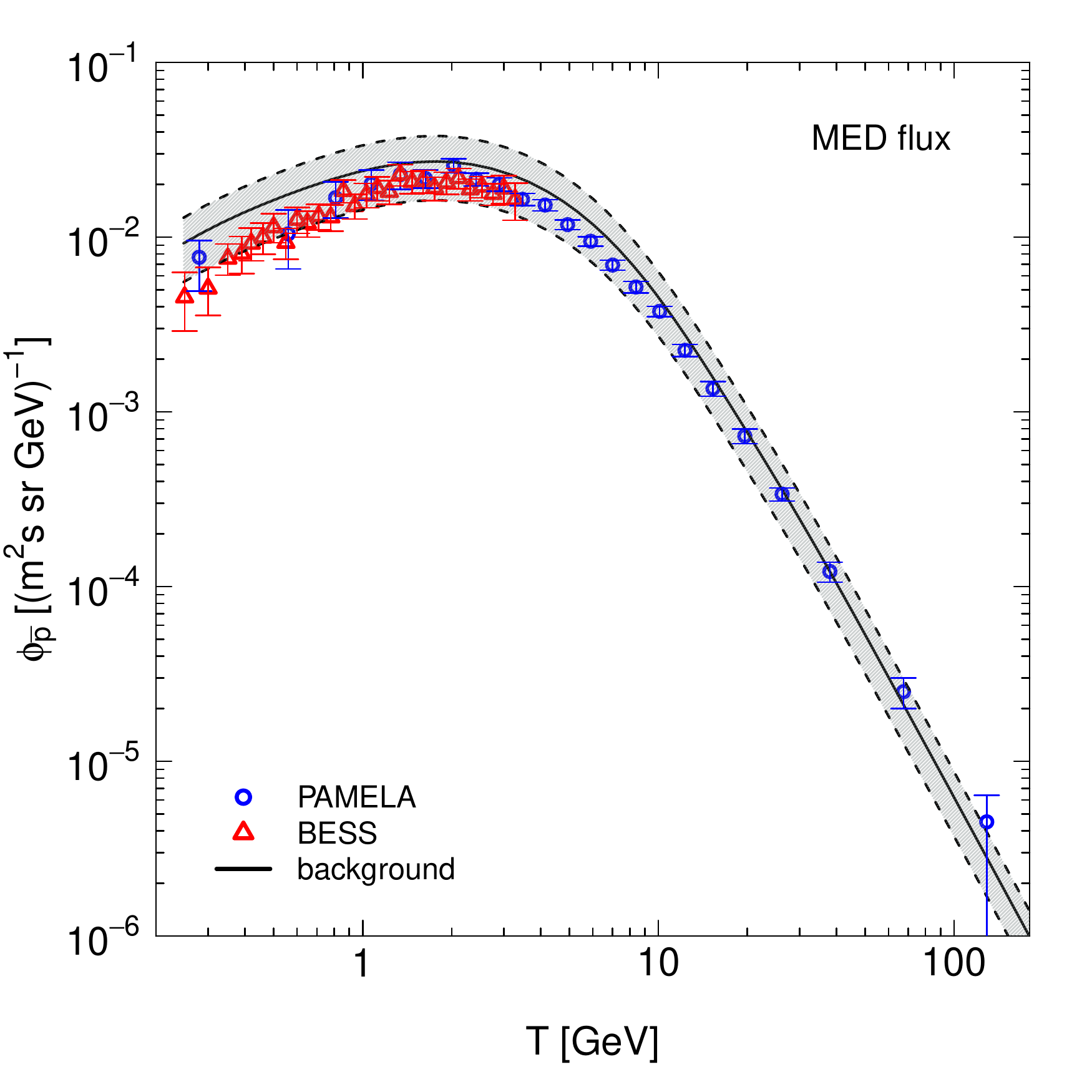}
\caption{Top-of-atmosphere antiproton flux $\Phi_{\pbar}$ as a function of the antiproton kinetic energy $T_{\bar p}$. Open circles (blue) data points refer to PAMELA  measurements 
\cite{Adriani:2010rc,Adriani:2012paa}.
Open triangles (red) data points refer to BESS-Polar \cite{Abe:2011nx}. The solid line shows the antiproton secondary production, propagated in the Galaxy with the MED set of transport parameters \cite{Donato:2008jk} and further propagated in the heliosphere with a charge-dependent solar modulation with propagation parameters 
$\alpha=20^\circ$, $\lambda=0.15$ AU and negative polarity. The band shows a (conservative) 40\%
theoretical uncertainties on the background calculation, mainly ascribable to nuclear-physics uncertainties in the production cross section and to uncertainties in the primary proton flux.}
\label{fig:TOA_spectra}
\end{figure}

The most recent, accurate and statistically significant datasets on cosmic antiprotons are currently provided by the space-borne PAMELA detector \cite{Adriani:2010rc,Adriani:2012paa} (in the kintic-energy interval between 90 MeV and 240 GeV) and by the balloon-borne BESS-Polar detector \cite{Abe:2011nx} (from 170 MeV to 3.5 GeV). The
top-of-atmosphere (TOA) fluxes are reported in Fig. \ref{fig:TOA_spectra}, together with the theoretical determination of the antiproton secondary production in the Galaxy obtained in Ref. \cite{Donato:2008jk}. The figure shows that secondary production is in good agreement with the data, and therefore 
additional (exotic) antiproton components, with dominant contribution in the 500 MeV to 50 GeV energy range, appear to be strongly constrained, unless significant modifications to the standard picture
of cosmic rays production and propagation are invoked.

The secondary background flux is the critical element in the derivation of bounds on exotic components,
including dark matter antiproton production. In Fig.  \ref{fig:TOA_spectra} we show the central estimate for the MED set of propagation parameters. Galactic propagation accounts for about a 20-30\% change  \cite{Donato:2008jk,Donato:2001ms} when the propagation model is varied inside the MIN/MED/MAX models described in Sect. \ref{sec:transport}.  This small uncertainties on the secondary antiproton flux reflects the relatively small uncertainties on the B/C data used for fixing the propagation model, and demonstrate that antiprotons and nuclei are suffering galactic propagation in a similar manner.

Fig.  \ref{fig:TOA_spectra} also shows a (conservative) uncertainty band. This theoretical uncertainty arises from uncertainties in the knowledge of the primary proton and helium fluxes, on the detailed mapping of the interstellar gas on which the primary protons impinge to produce the antiproton
background and most notably from uncertainties in the knowledge of the nuclear physics processes 
at the basis of the antiproton secondary production. These uncertainties are mostly related to the lack of updated data on the production cross sections at the center-of-mass energies relevant for low-energy cosmic rays studies.  While novel input on the primary cosmic
rays spectra will come from the forthcoming AMS-02 data, nuclear-physics cross sections are not
expected to experience significant improvements: a dedicated low-energy diffusion experiment would in
fact be a very useful tool for cosmic rays physics. In our calculations we will assume a total size of this uncertainty conservatively at the level of 40\% \cite{Donato:2001ms,Donato:2003xg}, and we will show in Section \ref{sec:AMS} that this theoretical uncertainty on the background flux is already important in the determination of the bounds on the DM properties, and will likely become a limiting factor in the ability to improve the bounds with the new, high statistics AMS-02 data.
Let us comment that in other recent analyses of antiproton data, like Ref. \cite{Cirelli:2013hv}, the uncertainty on the secondary flux has
been taken into account by allowing a free normalization and a free variation on the spectral index of the background flux: we instead assume the reference flux calculation of  \cite{Donato:2008jk,Donato:2001ms,Donato:2003xg}, obtained under physical assumptions, and allow for it a 40\% uncertainty. The
approach of using a physical reference flux is well-founded on the fact that the background flux is
calculated under the same physical assumptions used to determine the DM signal (same propagation
model) and using a physical model based on data for the determination of the secondary
production (primary proton and helium fluxes, gas distribution). 

\begin{figure}[t]
\centering
\includegraphics[width=0.45\textwidth]{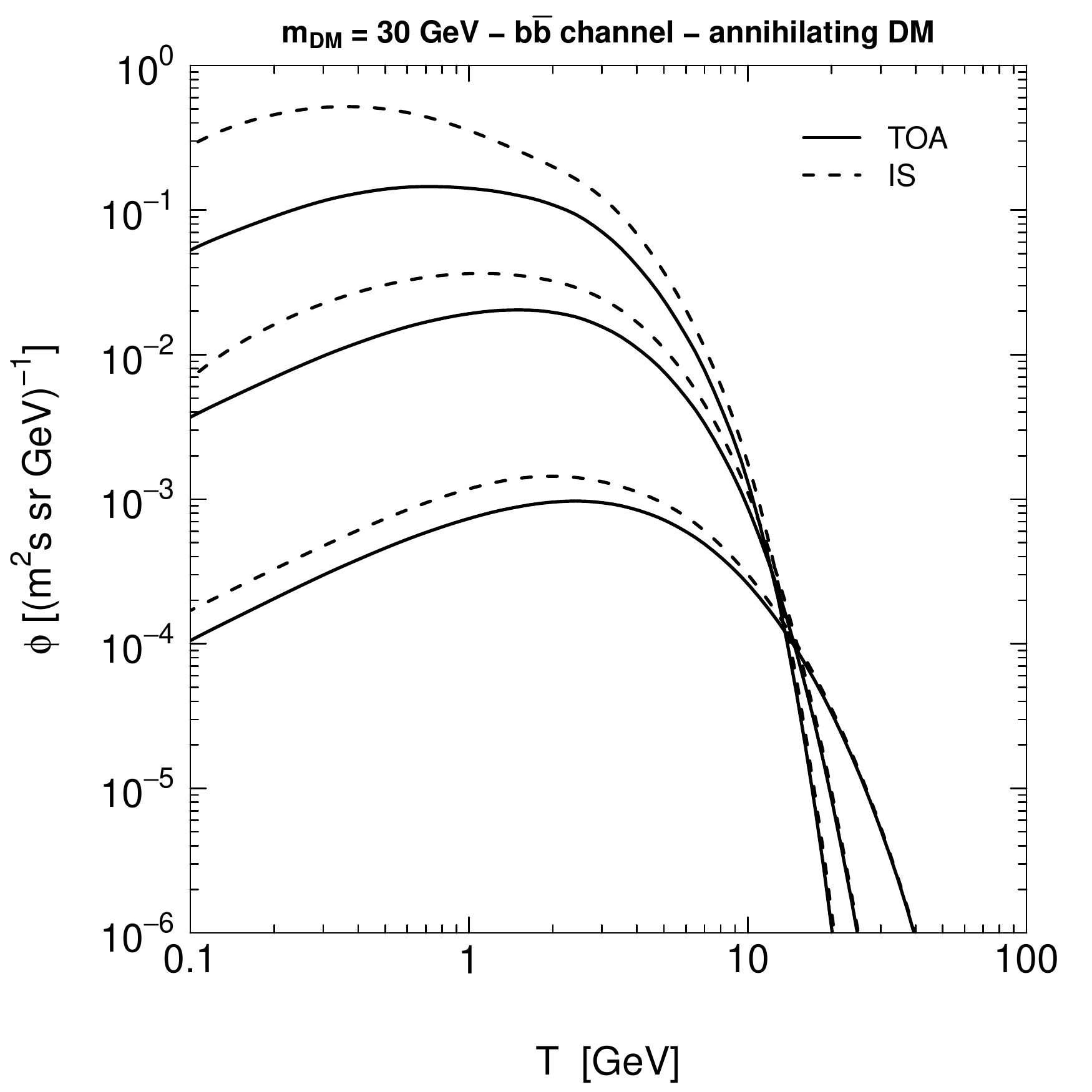}
\includegraphics[width=0.45\textwidth]{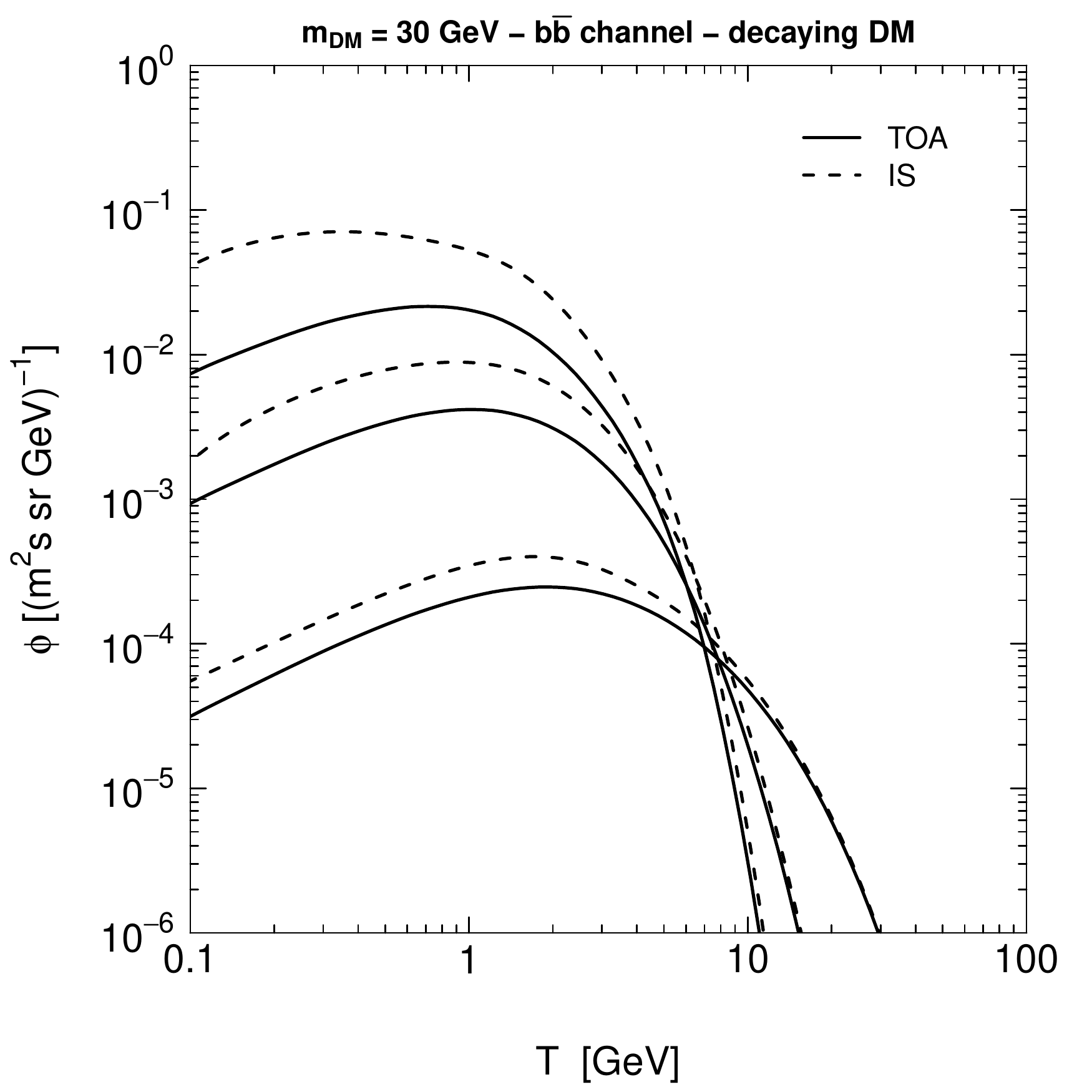}
\caption{Examples of antiproton spectra for DM annihilation (left panel) and DM decay (right panel) in the $\bar b b$ channel, $m_{\rm DM} = 30$ GeV and for a Einasto DM density profile. Dashed lines refer to the interstellar fluxes, solid lines to top-of-atmosphere fluxes, propagated in the heliosphere according to the modeling discussed in Section \ref{sec:solarmod} with a tilt angle $\alpha = 20 ^\circ$ and a mean-free-path $\lambda = 0.15$ AU. For definiteness, the annihilating case refers to a thermal cross section $\sigmav$, while the decaying case refers to $\tau=10^{28}$ s.
The upper/middle/lower set of curves refer to the MAX/MED/MIN sets for galactic transport.}
\label{fig:DMspectra}
\end{figure}

The antiproton flux from DM annihilation suffers a much larger variation from galactic transport modeling, as compared to the background. This variance can reach about a factor of 10 up (for the MAX model) or
down (for the MIN case), with some dependence on the antiproton energy \cite{Donato:2003xg}.

A specific example, which can help in guiding the discussion of the next Sections on the DM bounds,
is reported in Fig. \ref{fig:DMspectra}, where the antiproton spectra arising from a 30 GeV DM annihilating (left panel) and decaying (right panel) in the
$\bar b b$ channel, for a Einasto DM density profile, are reported. The figures show both the interstellar
fluxes (dashed lines) and the top-of-atmosphere (TOA) fluxes (solid lines). The latter have been obtained by propagating antiprotons in the heliosphere according to the modeling discussed in Section \ref{sec:solarmod} with a tilt angle $\alpha = 20 ^\circ$ and a mean-free-path $\lambda = 0.15$ AU. For definiteness, the annihilating case refers to a thermal cross section $\sigmav$, while the decaying case refers to $\tau=10^{28}$ s. The upper/middle/lower set of curves refer to the MAX/MED/MIN sets for galactic transport.

\begin{figure}[t]
\centering
\includegraphics[width=0.45\textwidth]{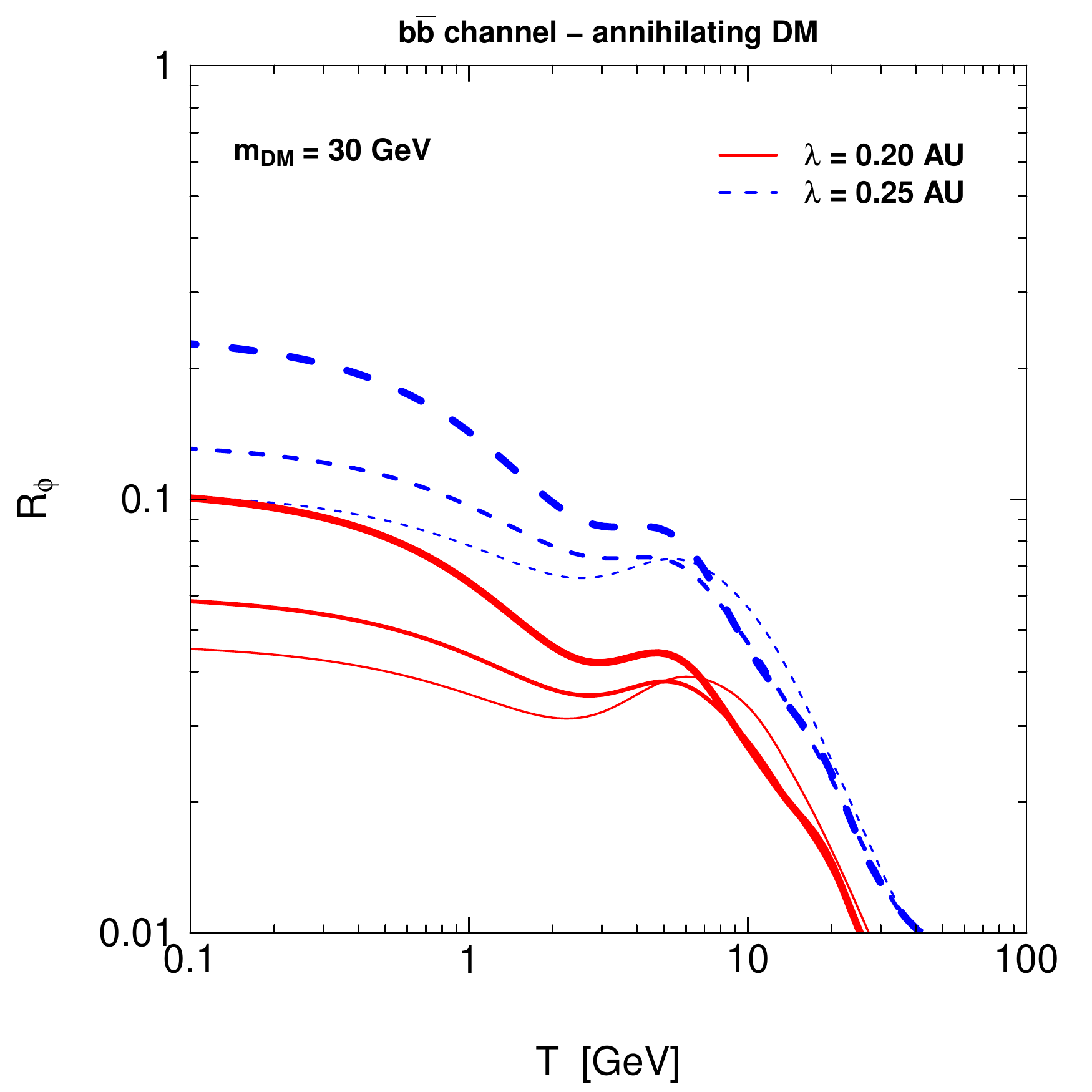}
\includegraphics[width=0.45\textwidth]{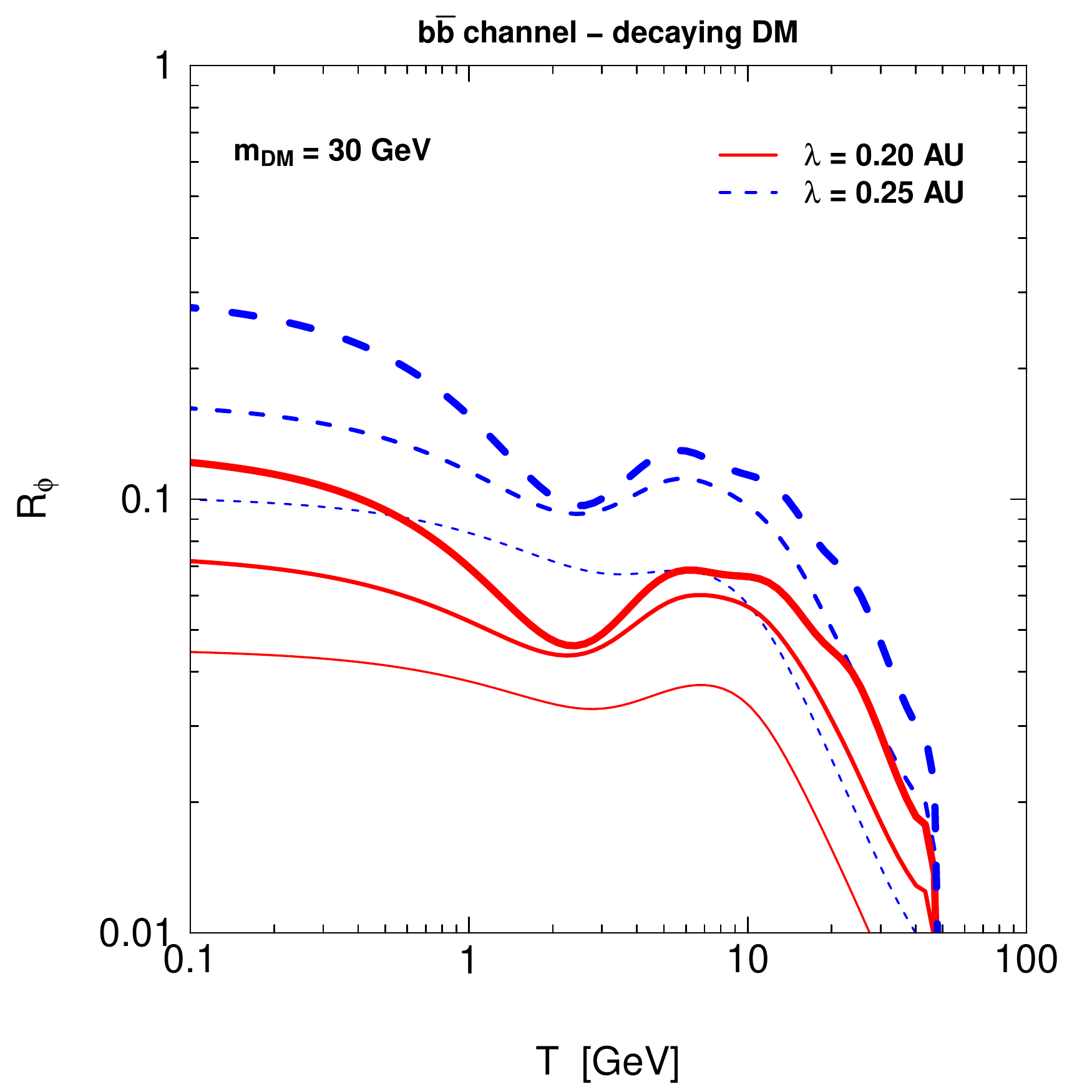}
\caption{Size of the effect induced on the TOA fluxes by solar modulation modeling. 
The figure shows the fractional variation of the
antiproton spectra $R_\phi = |1-\phi/\phi^{\rm ref}|$, where $\phi^{\rm ref}$ refers to
the TOA flux obtained with $\lambda = 0.15$ AU (i.e. the TOA fluxes shown in Fig. \ref{fig:DMspectra}).  In both panels, $\phi/\phi^{\rm ref}$ is always larger than 1. The left panel refers to annihilating DM, the right panel to decaying DM
and the production channel is $\bar b b$. Each panel has two sets of curves: solid lines
are obtained with $\lambda=0.20$ AU, dashed lines with $\lambda=0.25$ AU. For each set
of lines, the upper/median/lower curve refers to the MAX/MED/MIN set of galactic propagation
parameters.}
\label{fig:Pamela_solar_fractional_bb}
\end{figure}

The effect induced on the TOA fluxes by solar modulation modeling is shown in Fig. \ref{fig:Pamela_solar_fractional_bb}. The figure reports the fractional variation of the
antiproton spectra $R_\phi = |1-\phi/\phi^{\rm ref}|$, where $\phi^{\rm ref}$ refers to
the TOA flux obtained with $\lambda = 0.15$ AU (i.e. the TOA fluxes shown in Fig. \ref{fig:DMspectra}). The left panel refers to DM annihilating in the $\bar b b$ channel, the right panel to DM decaying in the same channel. These are representative cases: we have verified that a change in the annihilation channel does not alter significantly the results. Each panel has two sets of curves: solid lines are obtained with $\lambda=0.20$ AU, dashed lines with $\lambda=0.25$ AU. For each set
of lines, the upper/median/lower curve refers to the MAX/MED/MIN set of galactic propagation
parameters.  In both panels, $\phi/\phi^{\rm ref}$ is always larger than 1.

We notice that a change in solar modulation modeling has an  
impact which sizably differs depending on the interstellar flux, i.e. on the galactic transport model
at hand. In the MED case, the uncertainty on the TOA fluxes due to solar modulation is
maximal at lower kinetic energies, where it reaches the maximal size of 10\%  (15\% for decaying DM) in the energy range below 10 GeV. This maximal effect occurs for larger values of the mean free path $\lambda$. In the case of the MIN model, the largest uncertainties
are just around antiproton energies of 10 GeV, and they significantly decrease down to the few percent level at antiproton
kinetic energies below 1 GeV. In the MAX model, the effect is instead enhanced, and can reach
20\%-30\% for very low kinetic energies, slowly decreasing to 10\% at energies of 10 GeV. The origin
of this different impact of solar modeling is traced back to the different energy behavior of
the interstellar fluxes in the MIN/MED/MAX cases, as reported in Fig. \ref{fig:DMspectra}: larger confinement volumes allow for steeper interstellar fluxes in the 1-10 GeV kinetic-energy range
(the range which is more relevant in the determination of the TOA fluxes after solar modulation occurs)
and this therefore induces larger influence of solar modeling parameters in the the low-energy spectra
at the Earth. In the MIN case, the lower confinement volume produces interstellar fluxes which are
less steep in the few GeV range and this translates in less sensitivity of the TOA fluxes
on variation of solar modeling. As stated, a similar behavior is found for different production
channels.

While the most relevant source of variation on
the bounds arises from galactic propagation, a goal of our analysis is in fact to determine the impact
on the DM bounds arising from proper treatments of solar modulation. 
This is
a source of uncertainty which is independent from the one arising from galactic propagation: 
improvements in the galactic transport modeling, hopefully coming from the new cosmic-rays measurements of the AMS detector, will still leave open the issue of solar modulation. It is therefore
 a useful and novel piece of information to quantify these uncertainties. We will show that they can reach at most 50\% on the antiproton fluxes. The actual size
of the variation due to solar modulation modeling has a dependence on the signal production mechanism (annihilation vs. decay) and on the specific spectral features of the interstellar flux at the edge of the heliosphere (which is in turn determined by the specific galactic transport model). The impact of solar modulation uncertainties on the bounds on DM is therefore correlated to the galactic
transport modeling. 

Concurrently, solar modulation modeling allows us to use the whole available antiproton energy spectrum, including the low-energy PAMELA data, which are relevant to constrain light dark matter.
This explains a manifest difference in the bounds we derive here with those obtained in 
Ref. \cite{Cirelli:2013hv}: for DM masses below 50-80 GeV we obtain stringent bounds (coming
in fact from the low-energy part of the PAMELA dataset), while
Ref. \cite{Cirelli:2013hv} has much looser constraints in that mass range,
due to the adoption of PAMELA data only above 10 GeV.

\subsection{Statistical analysis}

For definiteness, we will present the bounds obtained from the PAMELA dataset \cite{Adriani:2010rc,Adriani:2012paa},
since it covers a wider energy range. Since PAMELA reports slightly larger fluxes
in the low-energy range, as compared to BESS-Polar, the derived bounds will be slightly more
conservative. We will use the PAMELA data in the rigidity range from 50 MV up to
180 GV, for which a statistically relevant measurement of the antiproton flux is available
(the highest-rigidity bin, which reaches 350 GV currently provides only an upper limit on the antiproton flux).

The bounds on the DM properties are reported as upper limits on the velocity averaged annihilation cross section $\sigmav$
(or lower limits in the case of the decay lifetime $\tau$) as a function of the DM mass $m_{DM}$, for the different annihilation/decay
channels which can produce antiprotons, and by assuming that the particle DM under study accounts for the whole DM in the Galaxy, regardless of the actual value of its annihilation cross section $\sigmav$ or decay lifetime $\tau$ (as it is customary). We adopt a rastering technique, where we determine bounds on $\sigmav$ (or $\tau$) at fixed values of the DM
mass $m_{DM}$. As a test statistic we employ a log-likelihood ratio $R$ defined as:
\begin{equation}
R=-2 \ln\left(\frac{\mathcal{L}}{\mathcal{L}_{bg}}\right)
\end{equation}
where ${\cal L}_{bg} = \prod_i f(E_i)_{bg}$ is the joint pdf of the background-only hypothesis
($i$ runs on the energy bins $E_i$) and ${\cal L}(\theta)_{bg+DM} = \prod_i f(E_i,\theta)_{bg+DM}$, where $\theta$ denotes either $\sigmav$ or $\tau$. By assuming independent energy
bins and gaussian pdfs, the test statistics is a chi-squared distribution with 1 degree of freedom, and we can set the bounds on the parameter $\theta$ by requiring that:
\begin{equation}
\Delta \chi^2 < n
\end{equation}
where $\Delta \chi ^2 = \chi ^2_{bg+DM} - \chi^2_{bg}$, with:
\begin{eqnarray}
\chi^2 _{bg} &=& \sum_i \frac{(\phi_i^{bg}-\phi_i^{exp})^2}{\sigma_{i,tot}^2} \\
\chi^2 _{bg+DM} &=& \sum_i \frac{(\phi^i_{bg+DM}-\phi^i_{exp})^2}{\sigma_{i,tot}^2}
\end{eqnarray} 
Let us comment that, as a consequence of experimental data being very well compatible with the background-only hypothesis, we have $\chi^2_{bg} \approx \chi^2_{\mathrm{best \,fit}}$.
We conservatively determine upper
[lower] bounds on $\sigmav$ [$\tau$] at a one-sided confidence level of 3$\sigma$ (i.e., CL = 99.86\%), which corresponds to $n=10.21$.
%
%

As discussed above, we allow theoretical uncertainties on the secondary background calculation
at the level of 40\%. The method we will adopt in the analysis is to
assume the errors $\sigma_{i,tot}$ to be composed by
two sources, which we add in quadrature:
\begin{equation}
\sigma_{i,tot}~=~\sqrt{\sigma_{i,exp}^2~+~\sigma_{i,theo}^2}
\label{eq:errorsum}
\end{equation}
where $\sigma_{i,theo} ~= ~0.4\times\phi_i^{bg}$,
as stated, and where the experimental errors $\sigma_{i,exp}$ contain both the 
statistical and systematic uncertainties, which we add linearly:
$\sigma_{i,exp} ~= ~\sigma_{i,stat} ~+~ \sigma_{i,sys}$\footnote{A linear sum between the statistical and systematic errors can be seen as an upper limit of a more usual quadrature sum, and it has been chosen in order to use the most conservative approach. Notice that our results are practically insensitive to this choice: in our analysis the theoretical uncertainty always largely dominates the experimental one.}. 

\begin{figure}[t]
\centering
\includegraphics[width=0.45\textwidth]{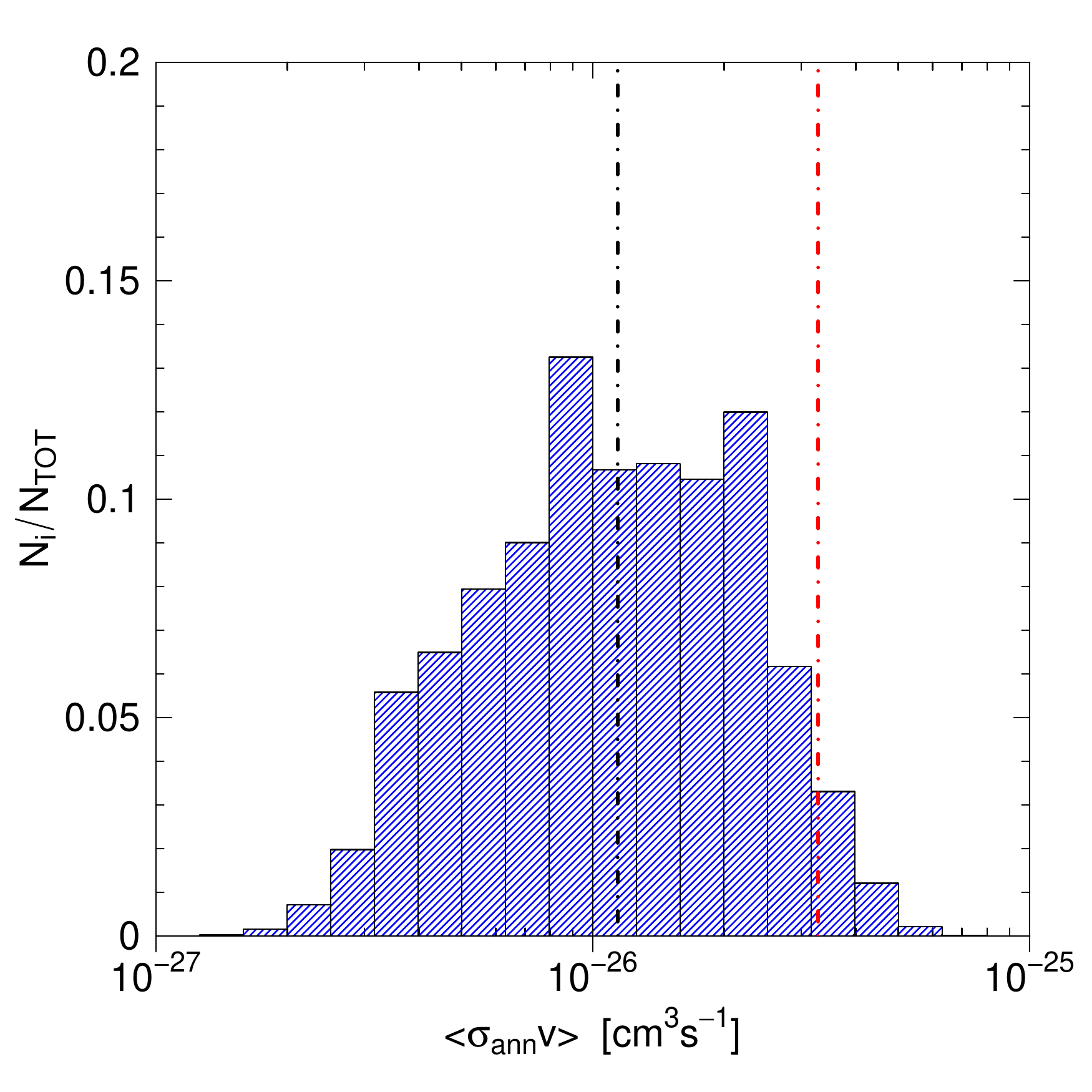}
\includegraphics[width=0.45\textwidth]{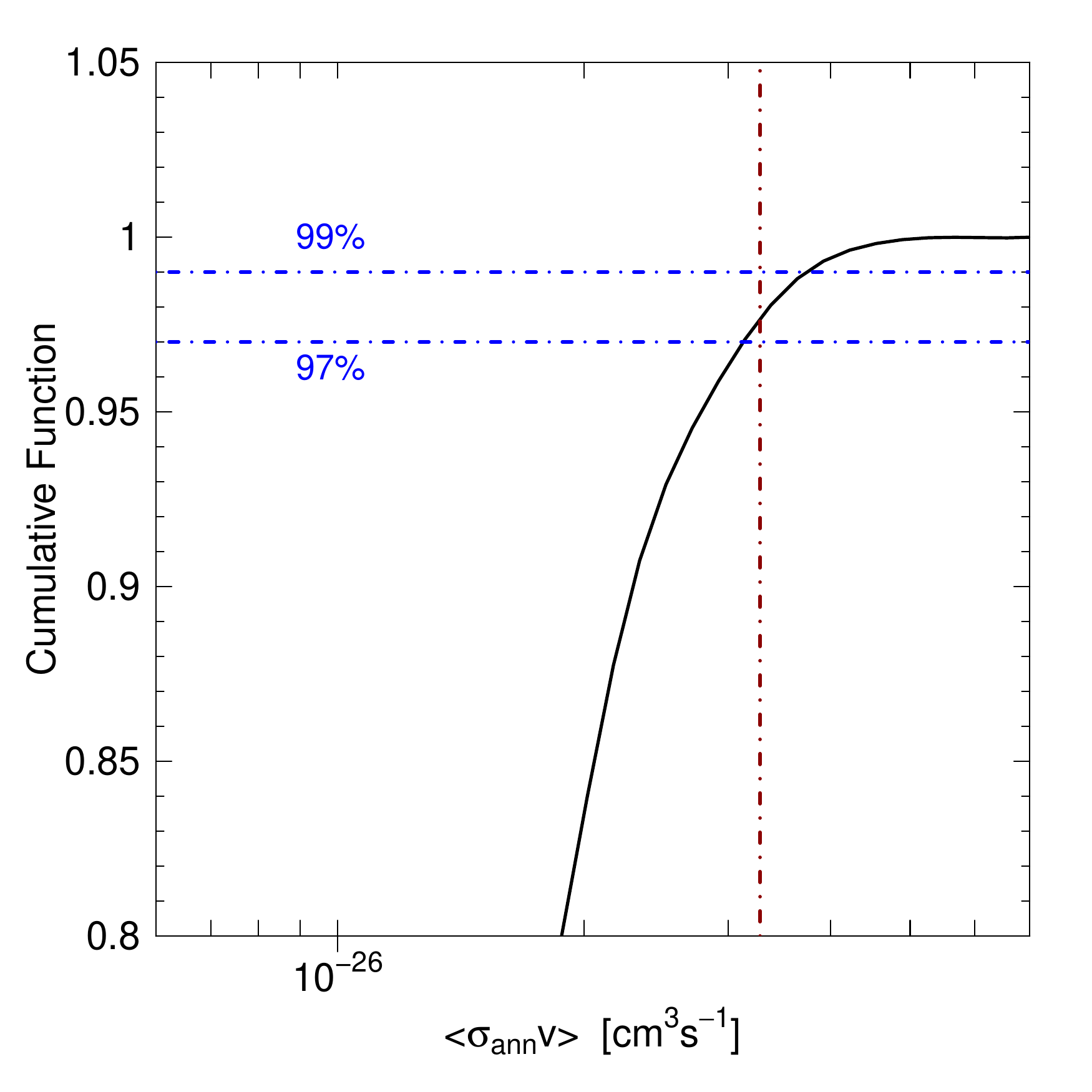}
\caption{{\sc Left panel}: Statistical distribution of  upper limits on the DM annihilation cross-section $\sigmav$ from the PAMELA data, obtained by using a Monte Carlo technique which takes into account
both experimental errors and theoretical uncertainties on the background-flux calculation. For definiteness, the plot shows the case of the $b\bar b$ annihilation channel for a DM mass of 50 GeV.
The vertical leftmost (black) line shows the bound obtained without considering the theoretical error
on the background calculation. The rightmost (red) line shows the bound obtained with the technique explained in Sec. \ref{sec:Bounds}, which has been
adopted in the present analysis. 
{\sc Right panel}: Cumulative function for the distribution of the bounds, obtained with
the  Monte Carlo technique. The vertical (red)  line corresponds to the bound obtained with the technique explained in Sec. \ref{sec:Bounds}, while the two horizontal red dot-dashed lines indicate the 97\% and the 99\% levels for this cumulative distribution.}
\label{fig:errors_distr}
\end{figure}

While this is a practical
way of including the theoretical uncertainties, a more proper and statistically correct way is to generate a large sample of realizations of the background flux, normally distributed around the background reference flux \cite{Donato:2008jk} and with a standard deviation of 40\%: for
each background realization, a bound is derived by using only $\sigma_{i,exp}$, and the ensuing distribution of the derived bounds on $\sigmav$ (or $\tau$) can be analyzed. This has been done in one
specific annihilation channel, in order to check the validity and the limitations of the method discussed above (which will be then adopted throughout). The left panel of Fig.~\ref{fig:errors_distr} shows the
statistical distribution of the $3\sigma$ upper bounds on $\sigmav$ obtained with $10^5$ statistical realizations of the background flux. The reference annihilation channel is 
$\bar b b$ and the bounds refer to a DM mass of 50 GeV. The mean value of the bounds is
$1.1 \times 10^{-26}$ cm$^{3}$ s$^{-1}$ (which corresponds to the upper limit obtained with the
reference background flux), with a relatively broad distribution. This means that nuclear uncertainties in the background calculation represent a critical element in the ability to
determine bounds on the DM properties (and on the possibility to detect a signal as well: with the upcoming AMS measurements, the dominant source of uncertainty will be in fact the theoretical one). The upper bound obtained with the technique discussed above is marked
by the rightmost (red) vertical line, which corresponds to the 98\% coverage of the
cumulative distribution of the bounds found in our Monte Carlo analysis, as is clear from the
right panel of Fig. \ref{fig:errors_distr}, where the cumulative distribution function is reported. This shows that by adding the theoretical uncertainty to the experimental errors, as
done in Eq. (\ref{eq:errorsum}), well (and conservatively) intercepts the actual fluctuations on the background calculations due to nuclear uncertainties.

\section{Constraints from PAMELA on the DM properties}
\label{sec:PAMELA}

\begin{figure}[t]
\centering
\includegraphics[width=0.45\textwidth]{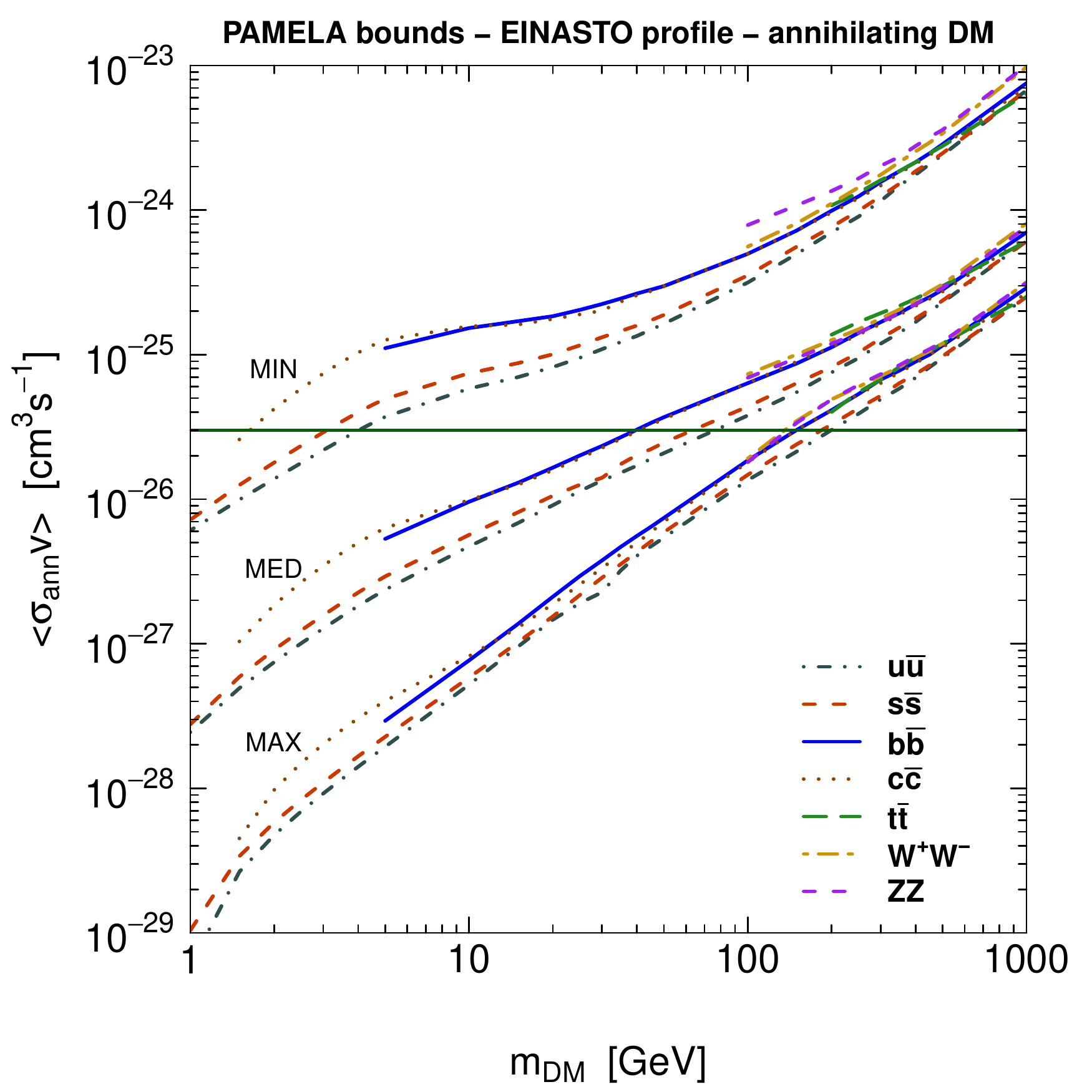}
\includegraphics[width=0.45\textwidth]{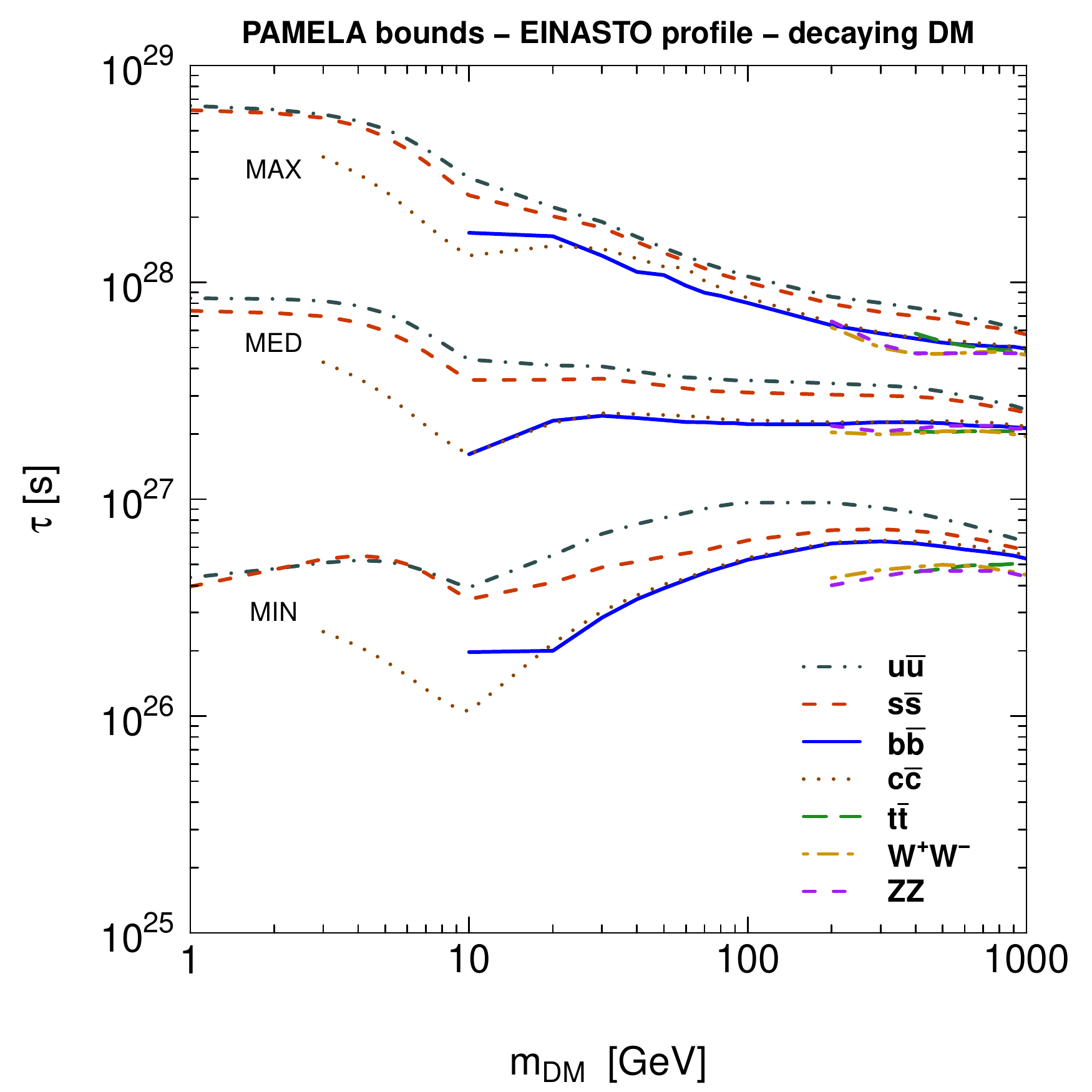}
\caption{Bounds as a function of the DM mass $m_{\rm DM}$  for the annihilating (left panel) and decaying (right panel) cases. For annihilating DM, the curves are upper bounds on the DM annihilation cross-section $\sigmav$. For  decaying DM, the curves are lower bounds on the DM decay lifetime $\tau$.
The  three sets of curves stand for the  MIN, MED and MAX sets of galactic propagation parameters, as reported in the labels, and are derived for an Einasto
DM profile. The different curves refer to different DM annihilation (or decay) channels.  The horizontal
line in the left panel denoted the ``thermal'' value $\sigmav = 3\times 10^{-26}$ cm$^3$ s$^{-1}$. For the annihilating case (left panel) the excluded region is above the relevant lines. For the
decaying case (right panel) the excluded region is below the lines. Lines start at the kinematical limit for signal production in the given channel: $m^{\rm lim}_{DM} = (m_i+m_j)$ for
for the annihilation channel $\chi+\bar\chi \rightarrow i+j$; $m^{\rm lim}_{DM} = (m_i+m_j)/2$ for
for the annihilation channel $\chi \rightarrow i+j$.}
\label{fig:Pamela_MED}
\end{figure}


Fig. \ref{fig:Pamela_MED} shows the bounds on the DM annihilation cross section $\sigmav$ 
(left panel) and on the DM lifetime $\tau$ (right panel), obtained from the PAMELA measurements for the various annihilation channels which can produce antiprotons ($u \bar{u}$, $s \bar{s}$, $c \bar{c}$, $b \bar{b}$, $t \bar{t}$, $ZZ$, $W^+W^-$). Fig. \ref{fig:Pamela_MED} refers to an
Einasto DM density profile for the galactic DM halo, and to the  MIN, MED and MAX sets of galactic propagation
parameters. For solar modulation we use a set of parameters compatible with the PAMELA  data-taking period: $\alpha = 20^\circ, \lambda = 0.15\;\mbox{AU}$. 
Clearly, for the annihilating case (left panel) the excluded region is above the relevant lines; for the decaying case (right panel) the excluded region is below the lines. Lines start at the kinematical limit for signal production in the given channel: $m^{\rm lim}_{DM} = (m_i+m_j)$ for
the annihilation channel $\chi+\bar\chi \rightarrow i+j$; $m^{\rm lim}_{DM} = (m_i+m_j)/2$ for the decay channel $\chi \rightarrow i+j$.

Fig. \ref{fig:Pamela_MED} shows that the bounds arising from antiproton measurements are actually quite stringent:  in the case of the MED propagation model, for light quarks, a thermal cross section is excluded for DM lighter
than about 80 GeV, while for heavier quarks (which produce smaller antiproton multiplicities)
the bound for thermal cross section is around 40 GeV. Light DM, below 10 GeV, is severely
bounded, both in the annihilating and decaying case. These bounds, obtained for the
central value of the allowed galactic-transport parameters set (the MED case)
are actually competitive with the limits obtained from gamma-rays measurements obtained with the Fermi-LAT detector, both from observations related to the extragalactic gamma-rays background and from observations of Milky Way satellites \cite{gammabound1,gammabound2,gammabound3,gammabound4,gammabound5,gammabound6,gammabound7,gammabound8}.

\begin{figure}[t]
\centering
\includegraphics[width=0.45\textwidth]{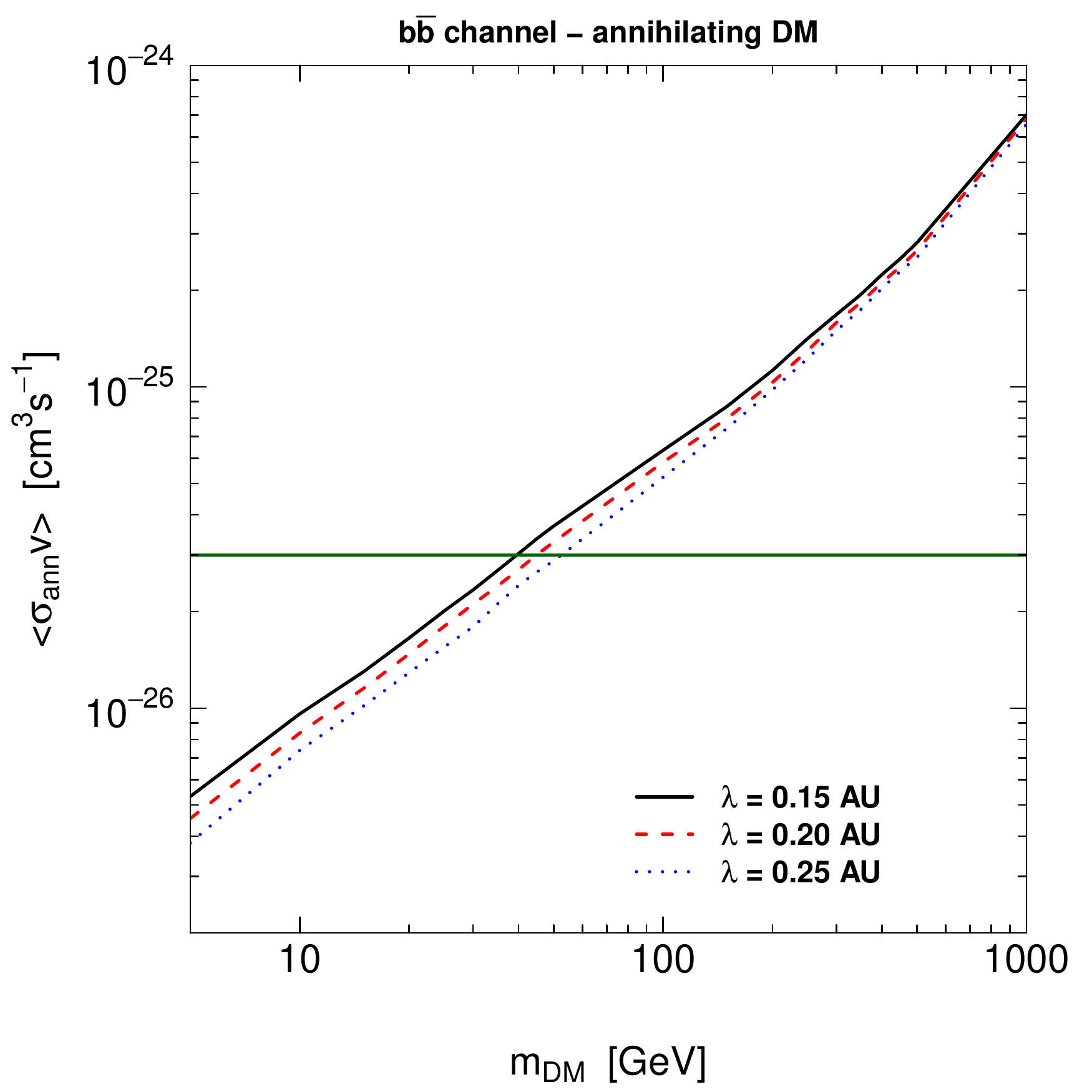}
\includegraphics[width=0.45\textwidth]{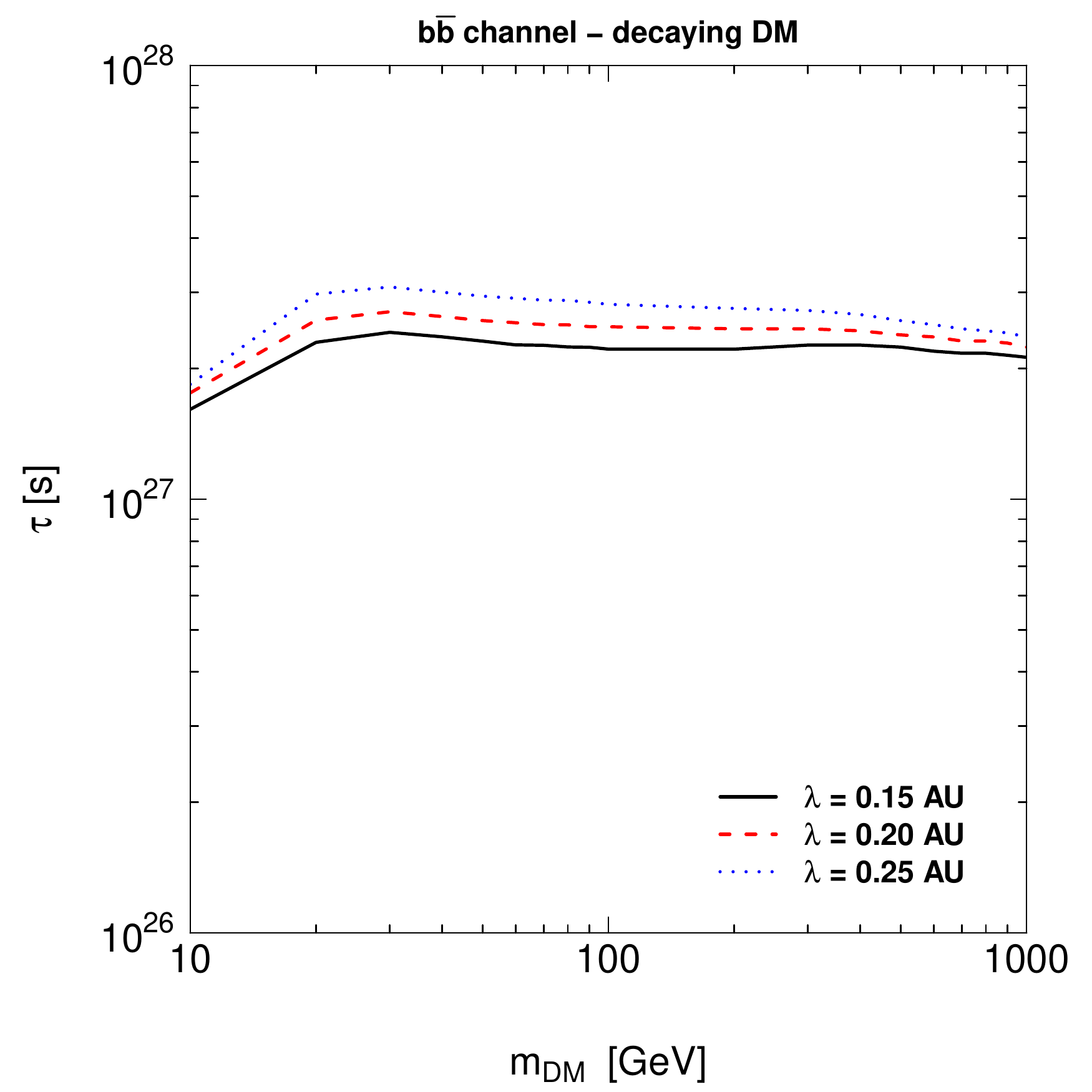}
\caption{Impact of the solar modulation modeling on the derived constraints for the annihilating (left panel) and decaying (right panel) case.  The different lines refer to a change in the mean free path
parameter $\lambda$: $\lambda=0.15$ AU (solid line), $\lambda=0.20$ AU (dashed line) and $\lambda=0.25$ AU (dotted line). For definiteness, the plots report the case of the $b\bar b$ annihilation (decay) channel.}
\label{fig:Pamela_solar}
\end{figure}


 Concerning the variation of the galactic transport modeling, as a result of the significant variation of the absolute
fluxes, as discussed above, the corresponding bounds are increased (decreased) by about an order
of magnitude for the MAX (MIN) set of propagation parameters, as compared to the MED case.
In the MIN case, thermal cross sections are excluded for DM masses below 3-4 GeV when
annihilation occurs into light quarks, while they are not constrained when DM annihilates into heavy quarks. In the case of the MAX set of parameters, very stringent bounds are present: for thermal cross
sections, all DM masses below 150 GeV are excluded. 
Concerning decaying DM, antiprotons set a lower bound on the lifetime of the DM particle at about $2\times 10^{27}$ s, which increases up to
10$^{28}$ s for DM masses of a few GeV and light-quarks production. These bounds are increased/decreased by about an order of magnitude for the MAX and MIN case. 

\begin{figure}[t]
\centering
\includegraphics[width=0.45\textwidth]{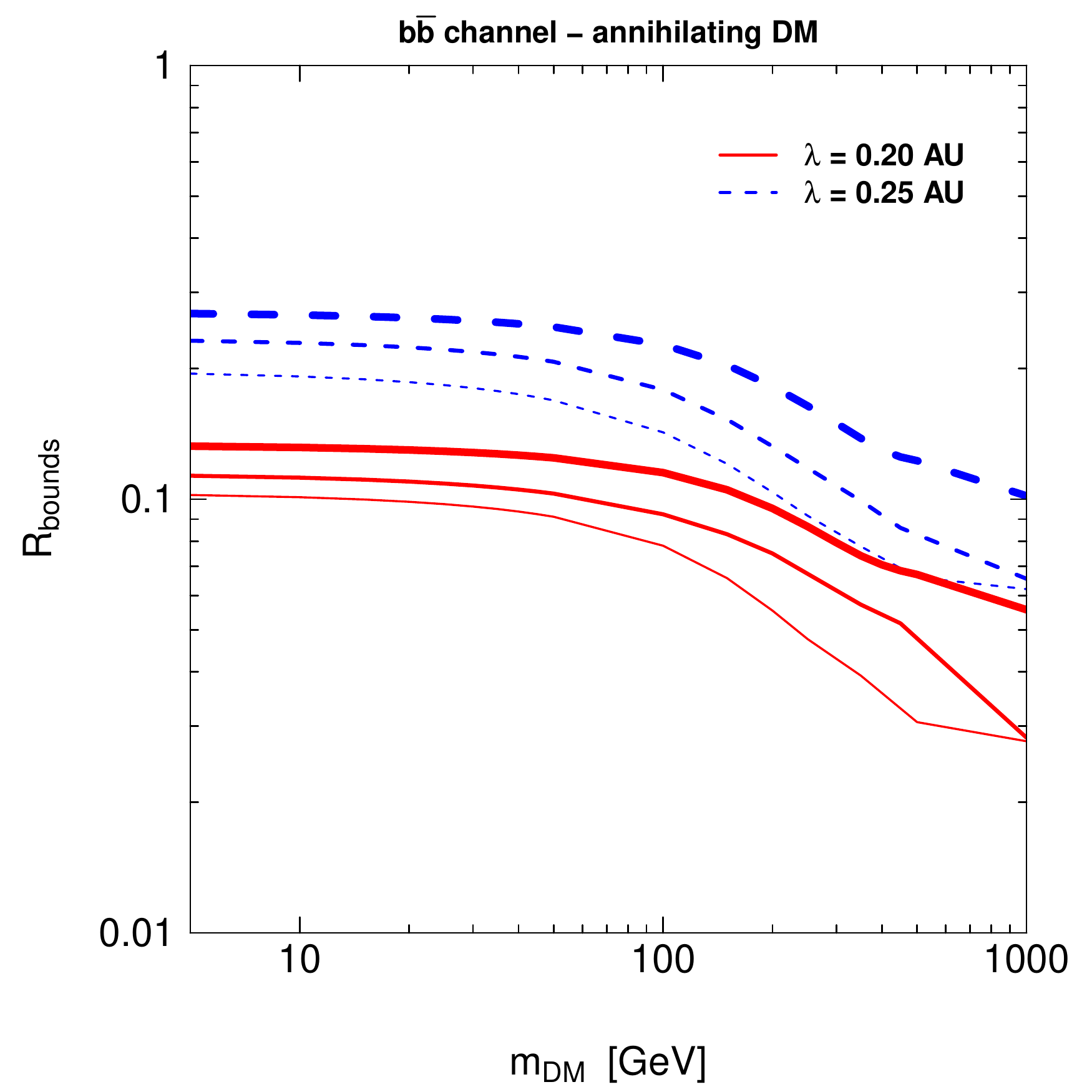}
\includegraphics[width=0.45\textwidth]{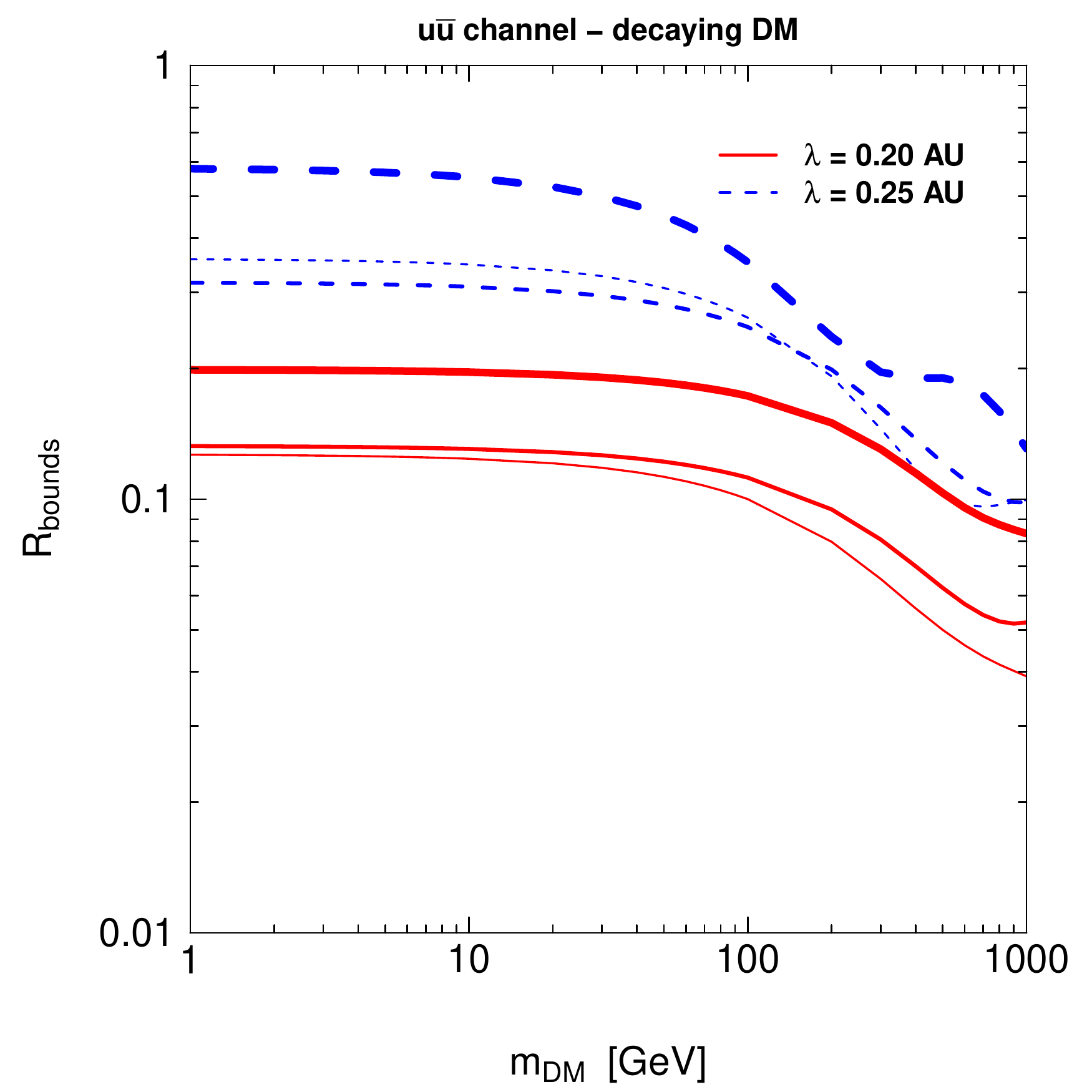}
\caption{
Impact of the solar modulation modeling on the derived constraints for DM annihilating into $\bar b b$ (left panel) and decaying into $\bar u u$ (right panel) case. The panels show the fractional variation 
of the
bounds $R_{\rm bounds} = |1-\theta_{\rm bound}/\theta_{\rm bound}^{\rm ref}|$ 
(where $\theta=\sigmav$ for annihilating DM, and  $\theta=\tau$) for decaying DM).
Each panel has two sets of curves: solid lines
are obtained with $\lambda=0.20$ AU, dashed lines with $\lambda=0.25$ AU (and the reference case
has $\lambda=0.15$ AU).
For each set
of lines, the upper/median/lower curve refers to the MAX/MED/MIN set of galactic propagation
parameters.
 For the annihilating case, $\theta_{\rm bound}/\theta_{\rm bound}^{\rm ref}<1$, while for
the decaying case $\theta_{\rm bound}/\theta_{\rm bound}^{\rm ref}>1$.}
\label{fig:Pamela_solar_fractional}
\end{figure}

The stringent bounds for DM lighter than about 50 GeV are mostly due to antiprotons arriving at
the top-of-atmosphere with energies below 10 GeV. Data at low kinetic energies therefore represent an important tool to probe DM: however, this is also the energy range where 
solar modulation is operative and therefore a proper treatment of cosmic-rays transport
in the heliosphere is important to determine the actual impact of antiproton measurements
in this DM mass sector. To this aim we have carefully modeled solar modulation transport with the techniques described in Sec. \ref{sec:solarmod}, and we have adopted different models compatible with the PAMELA data-taking period in order to quantify uncertainties on the bounds arising from solar modulation treatment. Results for the representative case of the $\bar b b$ channel are shown in Fig. \ref{fig:Pamela_solar},
where the bounds obtained with three different solar modulation models are reported
($\lambda=0.15$ AU, solid line; $\lambda=0.20$ AU, dashed line; $\lambda=0.25$ AU, dotted line).  Fig. \ref{fig:Pamela_solar} brings the information that in the MED annihilating case,
solar modulation modeling introduces an uncertainty of  40\% in the lower bound on the DM mass
for thermal cross sections: it moves from 40 GeV for $\lambda=0.15$  AU to 55 GeV for $\lambda=0.25$ AU.
 Fig. \ref{fig:Pamela_solar_fractional} shows the same information in terms of the fractional
variation of the bounds with respect to the result obtained for the reference model with $\lambda=0.15$ AU, i.e. $R_{\rm bounds} = |1-\sigmav_{\rm bound}/\sigmav_{\rm bound}^{\rm ref}|$  in the left panel and
$R_{\rm bounds} = |1-\tau_{\rm bound}/\tau_{\rm bound}^{\rm ref}|$  in the right panel. For illustrative purposes, the annihilating case refers to the $\bar b b$ production channel (representative of heavy quark production), the decaying
case $\bar u u$ (representative of light quark production).

From Fig. \ref{fig:Pamela_solar_fractional} we can see that, for galactic propagation set at the MED case, the largest variation of the bounds occurs, as expected, for light DM and is of the order of 
 25\% for annihilating DM and 40\% for decaying DM. This maximal variation occurs for solar models with larger
mean-free paths  $\lambda$ and is more relevant for light DM since in this case the bounds are
mostly induced by the lower energy bins of the PAMELA measurements. For DM masses around 100 GeV, the variation in the bounds due to solar modulation modeling  is still at the level of 10-15\%, and
decreases at a modest 5\% level when the DM mass approaches 1 TeV. 
Variation of the annihilation channel in terms of quark production produces
similar results.  Notice that for the annihilating case, $\theta_{\rm bound}/\theta_{\rm bound}^{\rm ref}<1$, while for
the decaying case $\theta_{\rm bound}/\theta_{\rm bound}^{\rm ref}>1$.

\begin{figure}[t]
\centering
\includegraphics[width=0.45\textwidth]{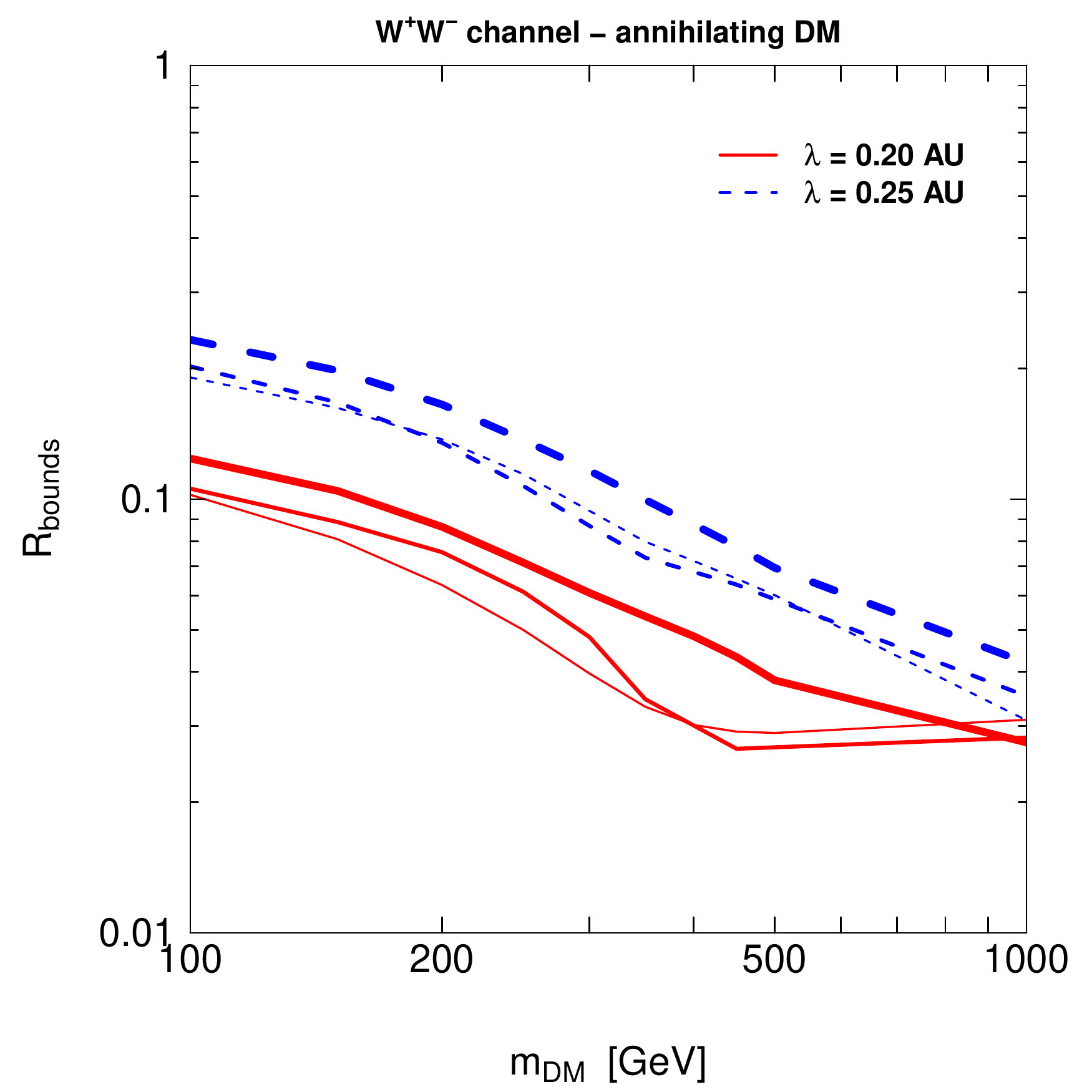}
\includegraphics[width=0.45\textwidth]{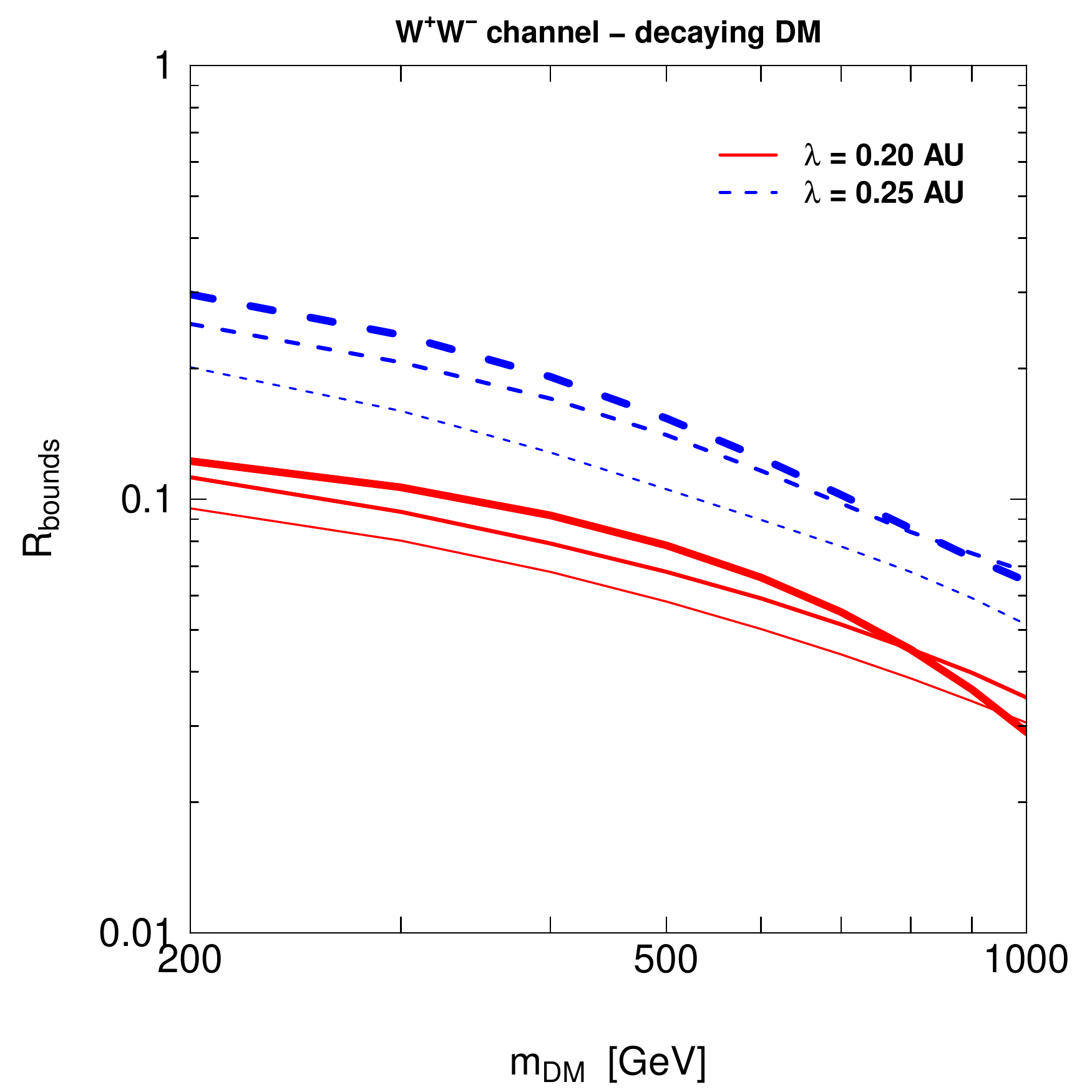}
\caption{The same as in Fig. \ref{fig:Pamela_solar_fractional_bb}, for the $\bar W^+W^-$ production channel.}
\label{fig:Pamela_solar_fractional_WW}
\end{figure}

Fig. \ref{fig:Pamela_solar_fractional_WW} shows the fractional variation $R_{\rm bounds}$ in the case of the $W^+W^-$ channel. Results are similar to the case of the  $\bar b b$ channel: for DM
masses of 100 GeV solar modulation modeling brings an uncertainty of the order of 20\% , which
steadily decreases to the few percent level for larger DM masses. In the case of gauge bosons production, the decrease in the uncertainty with the DM mass is steeper than in the case of quark production: this
is due to the fact that the gauge-boson channel is harder than the quark channel, and this implies that
the bounds on DM are coming from relatively larger energies, where solar modulation effects are
smaller.

We can therefore conclude that, in the case of the interstellar fluxes obtained with the MED galactic propagation, solar modulation modeling has an impact on the determination of antiproton
bounds, especially for DM masses lighter than 100 GeV, where the uncertainties can be seized to
be of the order of 20-40\%.


%

 Solar modulation modeling has an impact on the derived bounds which is less prominent than
what would be expected by just looking at the corresponding impact on the absolute fluxes, shown in 
Fig. \ref{fig:Pamela_solar_fractional_bb}. Fig. \ref{fig:Pamela_solar_fractional}, representative for the quark production channels, and Fig. \ref{fig:Pamela_solar_fractional_WW}, representative for the gauge-bosons production channels, show that the impact of a variation of solar modulation modeling remains
around 20-30\% for light annihilating DM and can reach 30-50\% for light decaying DM, regardless of the galactic transport model. The uncertainty is still of the same order of several tens of percent
for DM with a mass around 10 GeV, and decreases to the few percent level at 1 TeV. We notice that
in the case of the MAX galactic propagation, solar modulation uncertainties is always in excess of
10\% even for DM masses of 1 TeV, when the production channel is in terms of quarks.

 While these variations due to solar modulation modeling are not as large as those due to galactic
transport modeling, nevertheless they have a size that can influence the ability to set bounds
on the DM mass of annihilating DM which can reach at most 50\%, once a galactic transport model is
adopted, as discussed above. We can therefore conclude that uncertainty arising from solar modulation on the absolute fluxes is not dramatic,
although when this is transformed on impact on the mass bound of particle DM in a specific
model of galactic propagation, the influence is not completely negligible, and can change the bound
for a particle with themal $\sigmav$ by 15 GeV, as can be seen in the left panel of Fig. \ref{fig:Pamela_solar}.

\begin{figure}[t]
\centering
\includegraphics[width=0.45\textwidth]{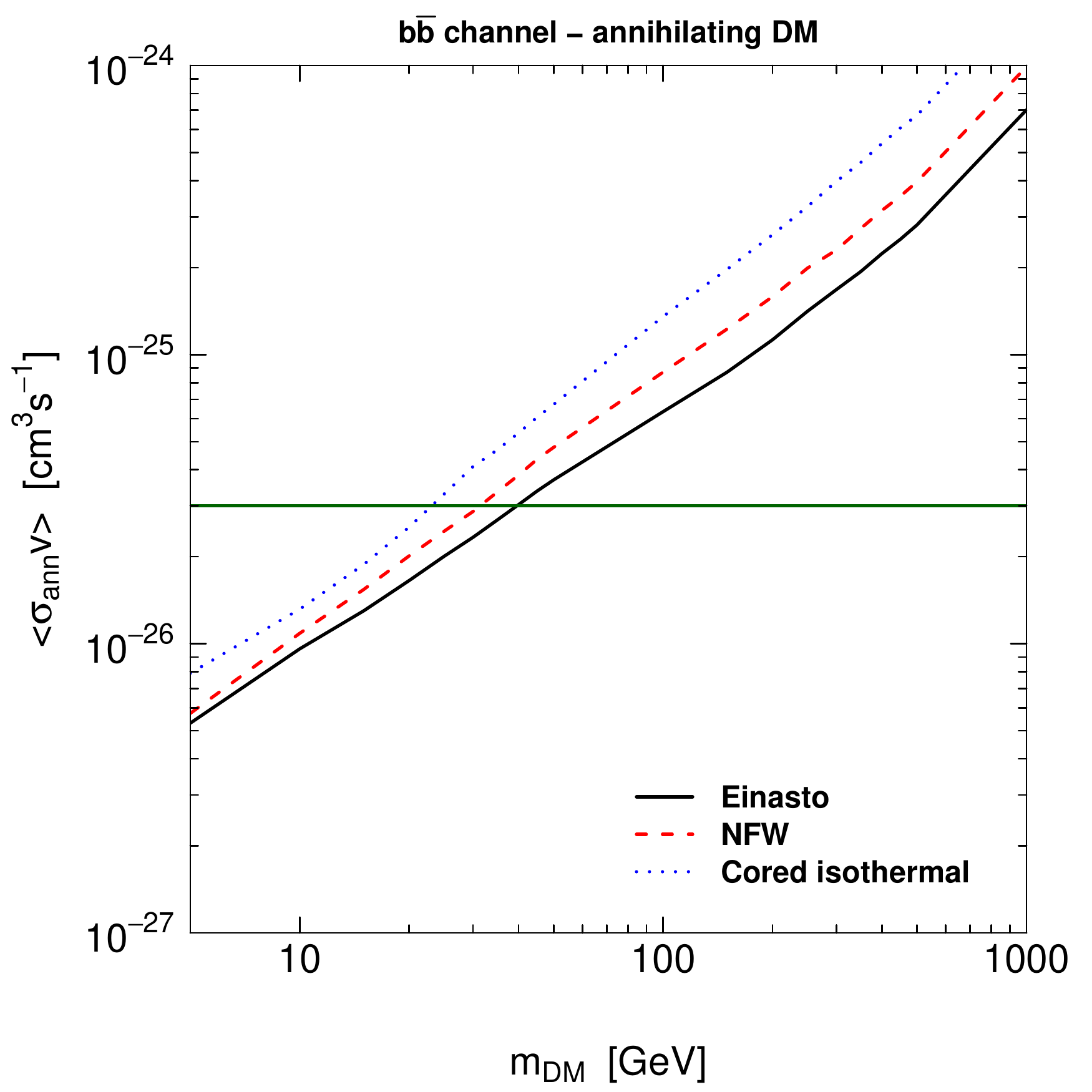}
\includegraphics[width=0.45\textwidth]{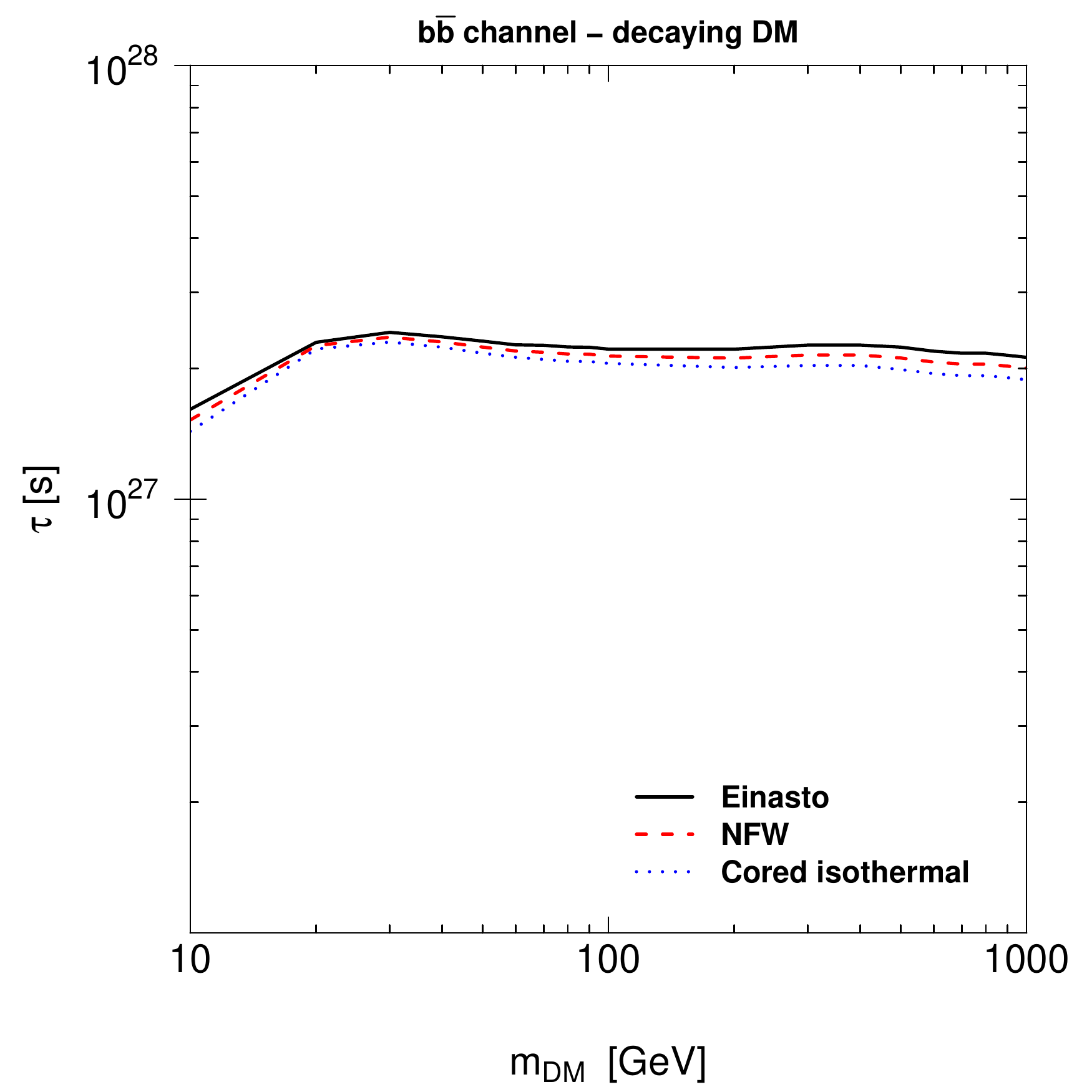}
\caption{Impact of the DM density profile on the derived constraints for the annihilating (left panel) and decaying (right panel) case. The different lines refer to the Einasto (solid line), NFW (dashed line) and cored isothermal (dotted line) profiles. For definiteness, the plots report the case of the $b\bar b$ annihilation (decay) channel.} 
\label{fig:Pamela_profile}
\end{figure}


\section{Prospects for AMS-02}
\label{sec:AMS}


In this Section we derive prospects for a 13 years data-taking period of the Alpha Magnetic Spectrometer (AMS-02), which was deployed on the International Space Station in May 2011. AMS-02 is an experiment designed to give precision measurements of a wide number of cosmic-rays species, including antiprotons. This will allow possible improvements in the determination of antiproton bounds
on DM: larger statistics and reduced systematics on the antiproton spectrum; improved data on
the primary flux, which could help in reducing the uncertainty on the theoretical
determination of the secondary antiproton background; improved data on cosmic rays nuclei, which could be instrumental to reduce the galactic transport uncertainties; large statistics data over a long
exposure time on a large number of cosmic rays species (hadronic and leptonic), which could help in better shaping transport modeling in the heliosphere. On the other hand, the extension of latitudes covered
by the International Space Station trajectory will limit the minimal accessible energies, due to
the geomagnetic cutoff. 

\begin{figure}[t]
\centering
\includegraphics[width=0.45\textwidth]{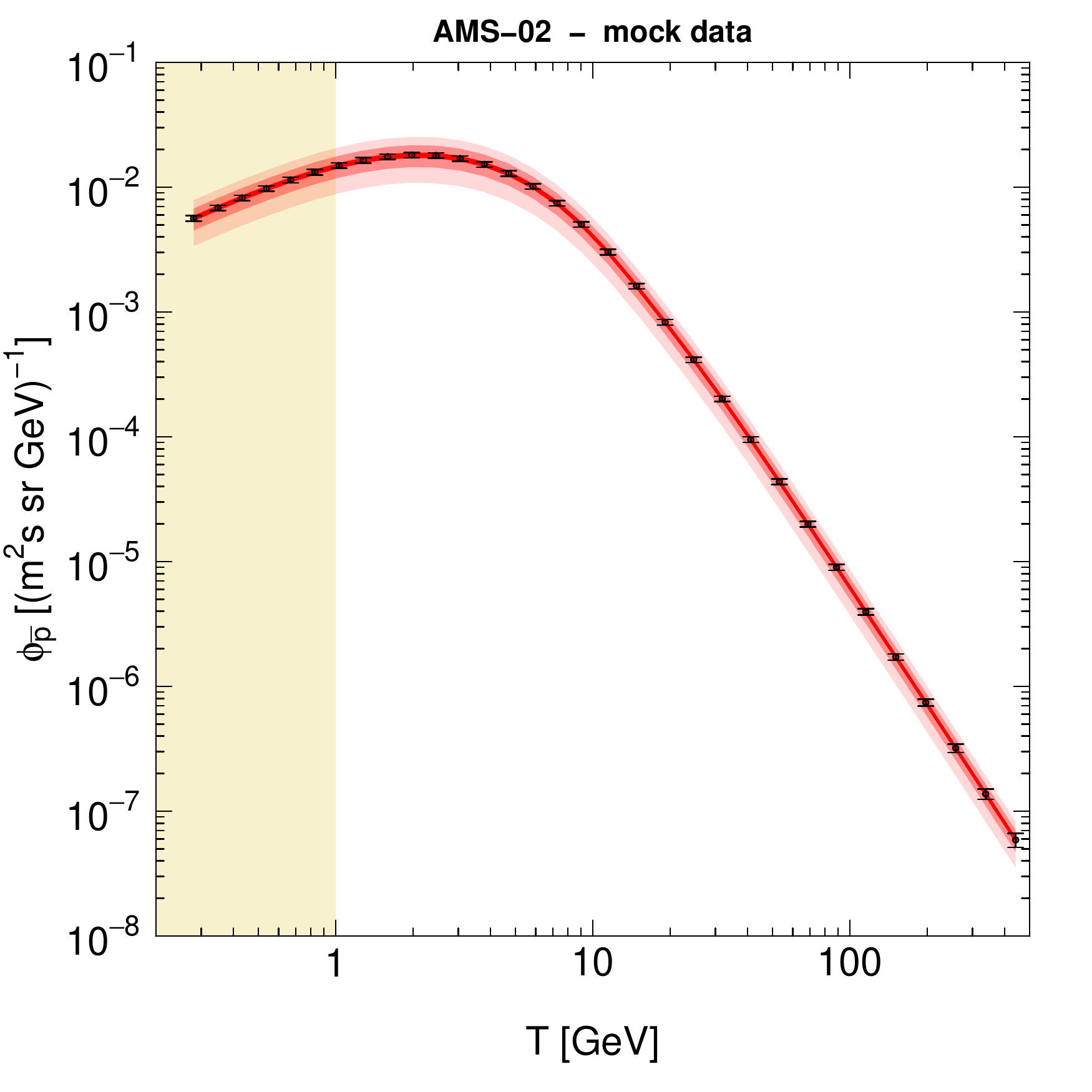}
\caption{Mock data for the AMS mission, used in the analysis for the AMS projected sensitivity.
The mock data are generated from the central value of the antiproton theoretical background of Fig. \ref{fig:TOA_spectra}. The three shaded bands around the mock data refer to a 40\%, 20\% and 5\% uncertainty
around the theoretical expectation. The vertical band for $T<1$ GeV denotes the energy
range not used in the analysis, because of the impact of the geomagnetic cutoff.} 
\label{fig:AMS_mock}
\end{figure}

We perform the analysis of the prospects for AMS-02 by generating mock data according to the
AMS-02 specifications and by adopting on the mock data the same analysis technique described in Sec. \ref{sec:Bounds}, and used in Sec. \ref{sec:PAMELA} for the analysis of the PAMELA data. The mock data are generated under the hypothesis of the presence of background only, for which we adopt the theoretical estimate of  Ref. \cite{Donato:2008jk}, i.e. the median curve of Fig. \ref{fig:TOA_spectra}.

Concerning solar modulation, since the
AMS-02 operational period will likely be very long (we consider a duration
from 2011 to 2024) and will cover more than one solar cycle, we subdivide the 
data-taking period in three phases, for which we adopt the following solar modeling: 
\begin{itemize}
\item{{\bf phase 1} (2011-2013): negative polarity of the SMF and solar activity close to the maximum; for this phase we will consider a tilt angle $\alpha=60^{\circ}$ }
\item{{\bf phase 2} (2013-half 2015 and 2021-2024): positive polarity of the SMF and solar activity nearly maximal, which again is compatible with a tilt angle of $\alpha=60^{\circ}$}
\item{{\bf phase 3} (half 2015-2021): positive polarity of the SMF and a solar activity nearly minimal , for which we use a tilt angle of $\alpha=20^{\circ}$)}  
\end{itemize}
 \begin{figure}[t]
\centering
\includegraphics[width=0.45\textwidth]{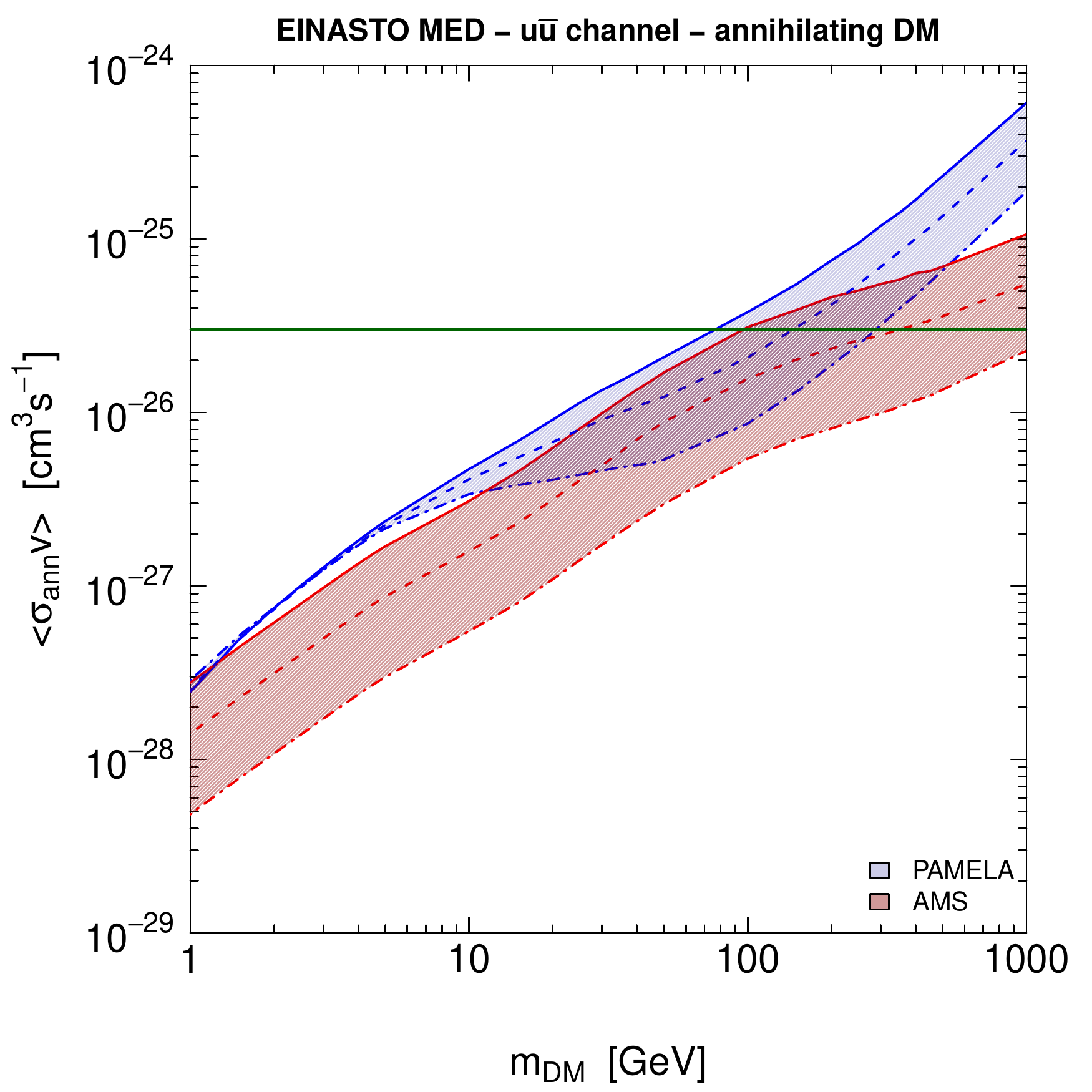}
\includegraphics[width=0.45\textwidth]{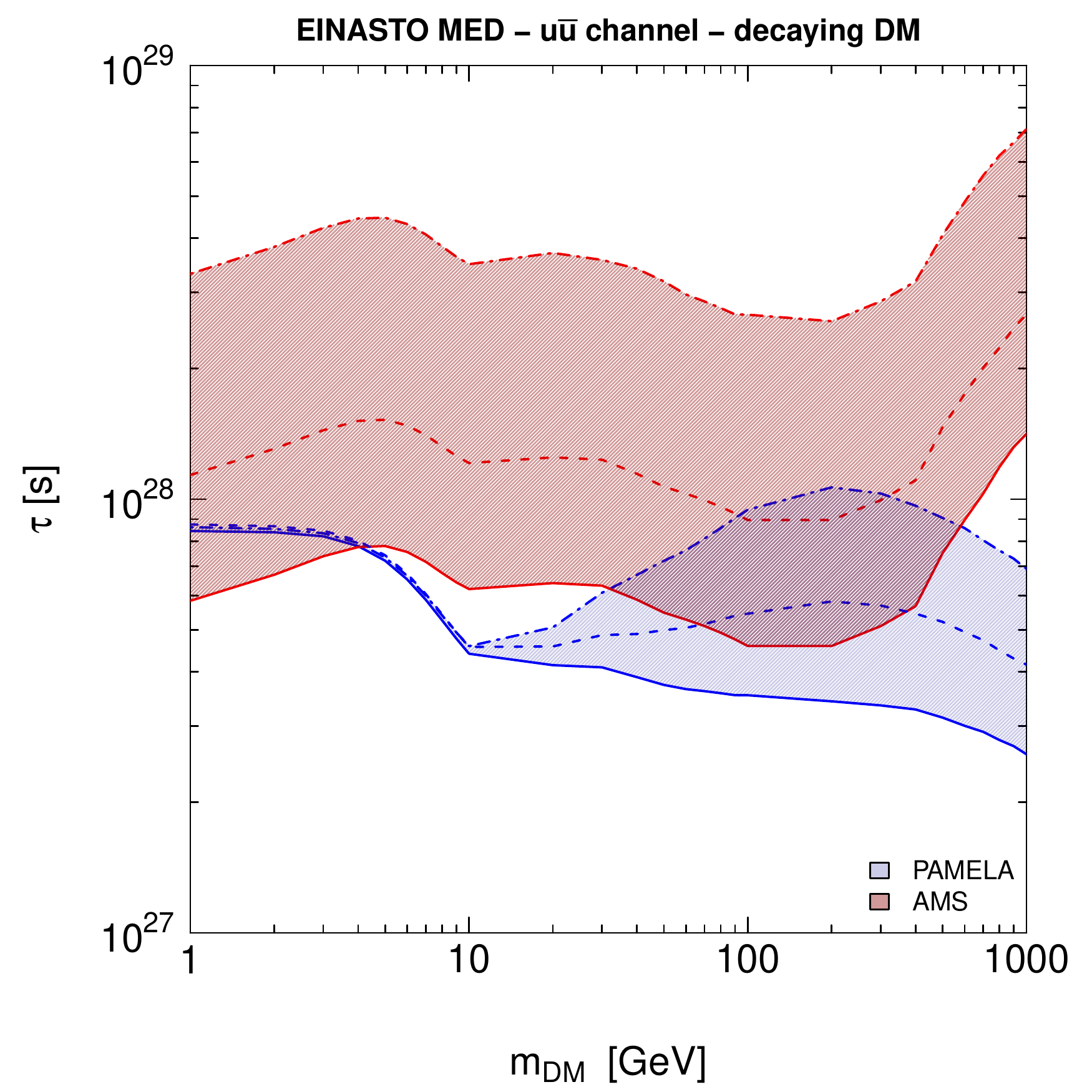}
\caption{Projected sensitivity for AMS-02, for annihilating (left panel) and decaying (right panel) DM,
compared to the current bounds from PAMELA. The representative case reported here refers to DM annihilation/decay into $u\bar u$, an Einasto density profile and the MED set
of  propagation parameters in the Galaxy. In the derivation of these bounds, it
has been assumed a low-energy threshold (due to the geomagnetic cut-off) for AMS-02 of $T^{\rm min}_{\bar p} = 1$ GeV.
Each set of curves (in the left panel the ``upper'' blue band refers to PAMELA, the ``lower'' red band refers to AMS-02; the reverse occurs in the right panel: the `'lower'' blue band refers to PAMELA, the ``upper'' red band refers to AMS-02) show the current PAMELA bound or the projected AMS-02
sensitivity, under three different assumptions on the size of the theoretical uncertainties on the secondary
antiproton production: solid, dashed and dot-dashed lines refer to 40\%, 20\% and 5\%, respectively. The solid lines for PAMELA reproduce the bounds
 reported in Fig. \ref{fig:Pamela_MED}. The horizontal (green)
line in the left panel denotes the ``thermal'' value $\sigmav = 3\times 10^{-26}$ cm$^3$ s$^{-1}$.} 
\label{fig:AMS_uu}
\end{figure}

We determine the energy binning of the mock data by first determining the AMS-02 resolution in
the energy range of interest (which is here below 500 GeV). This is directly derived from the rigidity resolution which, following Ref., \cite{Cirelli:2013hv} can be parametrized as: 
\begin{equation}
\frac{\Delta \cal{R}}{\cal{R}} = 0.00042~\times~{\cal{R}} ~+~ 0.01
\end{equation}
From the rigidity resolution, the energy resolution is directly obtained as: 
\begin{equation}
\frac{\Delta T}{T} = \frac{T+2m_p}{T+m_p}~\frac{\Delta {\cal{R}}}{{\cal{R}}}
\end{equation}
Then, we require that mock-data bins are comparable in size to the energy resolution: in agreement with Ref. \cite{Evoli:2011id}, we adopt 10 bins per energy decade. In the energy bin with a central energy value $T_i$ and a width $\Delta T_i$, the number of expected antiproton events is then given by:
 \begin{equation}
 N~=~\epsilon a(T_i)\phi(T_i)\Delta T_i\Delta t
 \end{equation}
where $\epsilon$ denotes the efficiency (we assume $\epsilon=1$, for definiteness), $\Delta t$ is the length of the data taking period,  $a(T_i)$ denotes the energy-dependent acceptance, which we
assume as in Ref. \cite{Cirelli:2013hv}:  for $T~<~11$ GeV we assume $a(T)~=~0.147$ m$^2$ sr,  
for larger kinetic energies we derive an energy dependence by fitting the curve in Fig. 8 of Ref. \cite{AMS_acc}. Finally, we assume that the statistical error of the mock data in each energy bin is poissonian, and we allow for a 5\% systematic uncertainty. The generated AMS mock data,
together with the theoretical uncertainty bands of 40\%, 20\% and 5\% sizes, are reported
in Fig. \ref{fig:AMS_mock}.

Due to geomagnetic effects, the efficiency $\epsilon$ will drop 
starting from energies of about 30 GeV, down to sub-GeV energies where the detection efficiencies
(or, alternatively, the effective area) will be reduced to few percent of its nominal value \cite{geo_battiston}. For this reason,
we include in the analysis of AMS mock data only the energy range above
$T_{\rm min} = 1$ GeV.

\begin{figure}[t]
\centering
\includegraphics[width=0.45\textwidth]{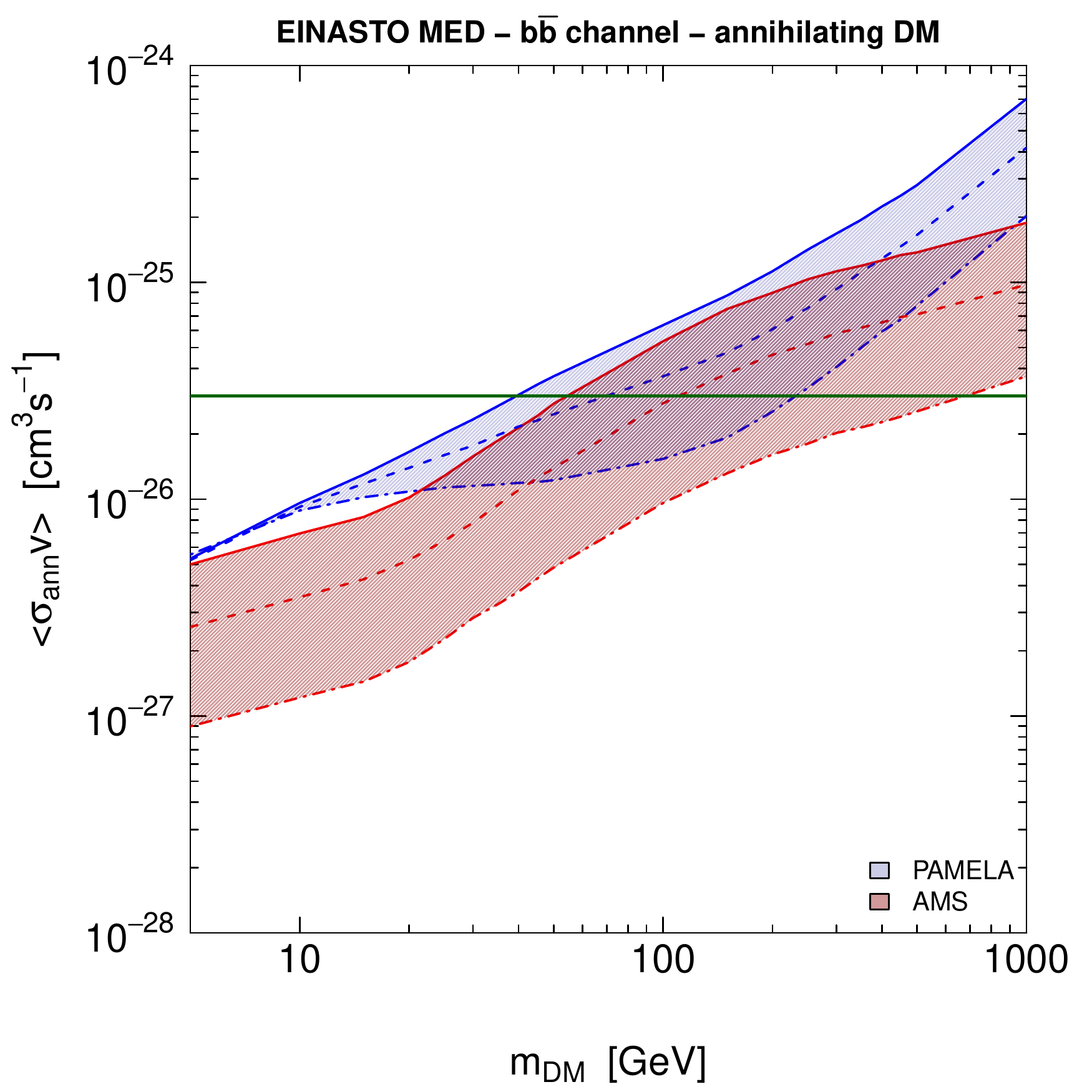}
\includegraphics[width=0.45\textwidth]{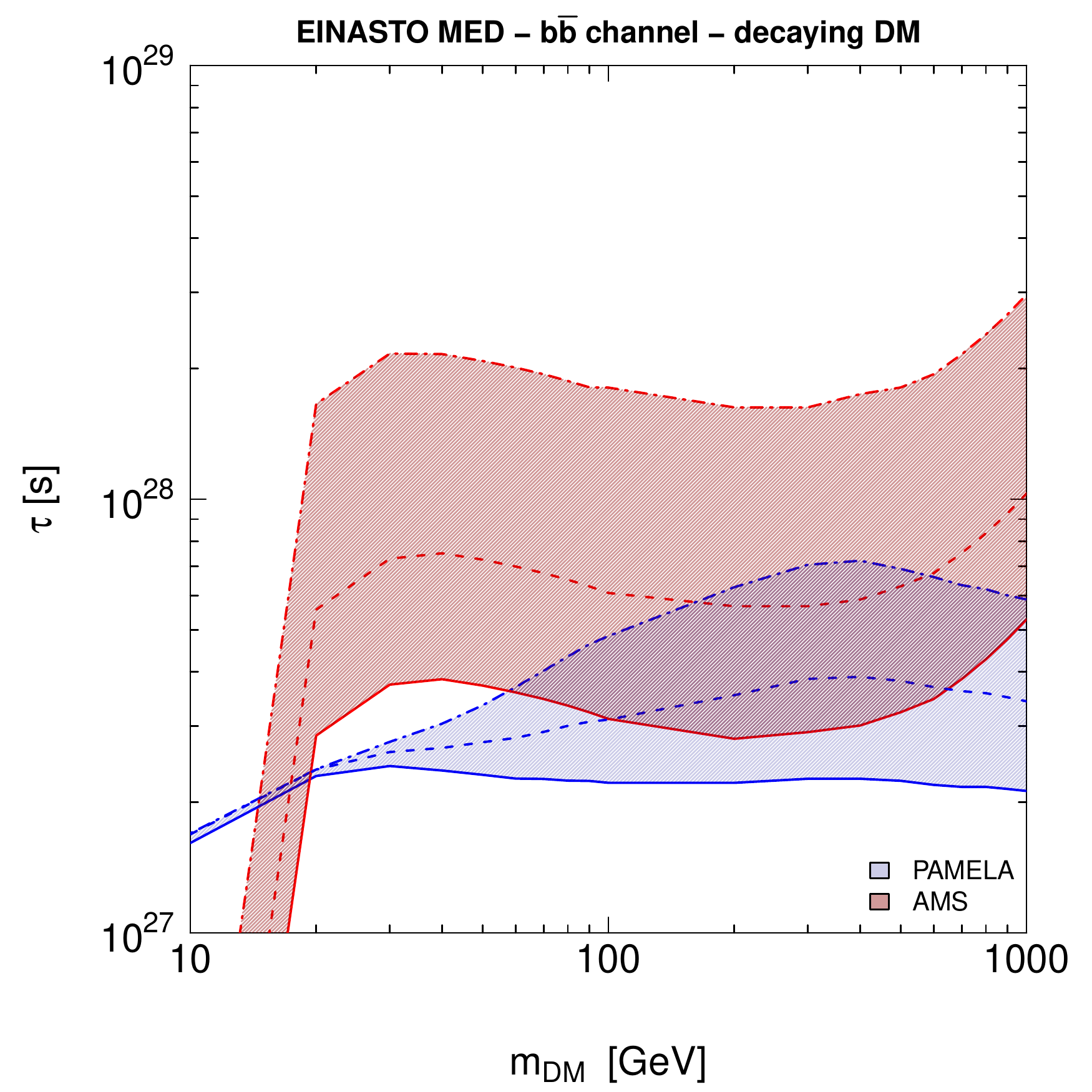}
\caption{The same as in Fig. \ref{fig:AMS_uu}, for the $b \bar b$ annihilation/decay channel.}
\label{fig:AMS_bb}
\end{figure}

\begin{figure}[t]
\centering
\includegraphics[width=0.45\textwidth]{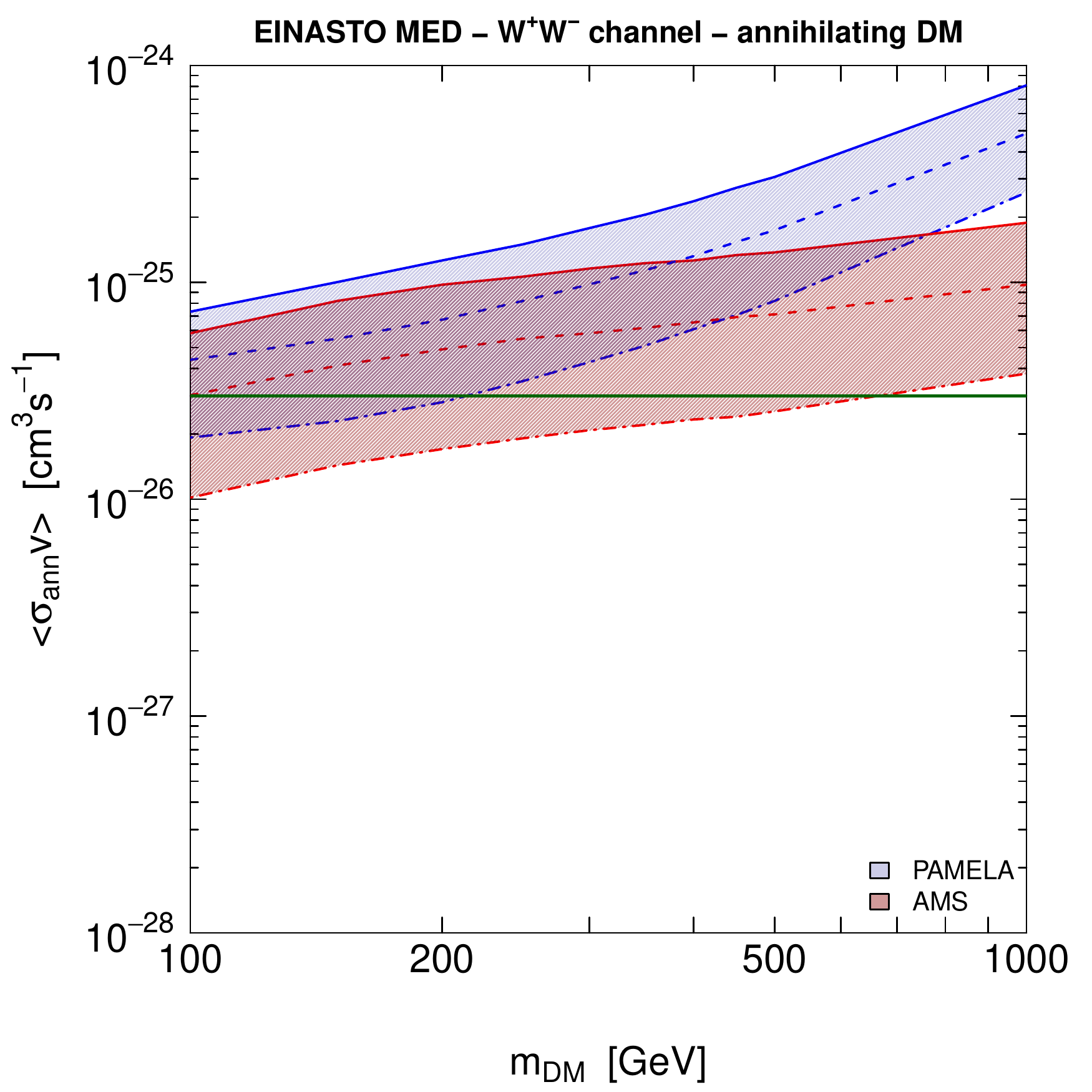}
\includegraphics[width=0.45\textwidth]{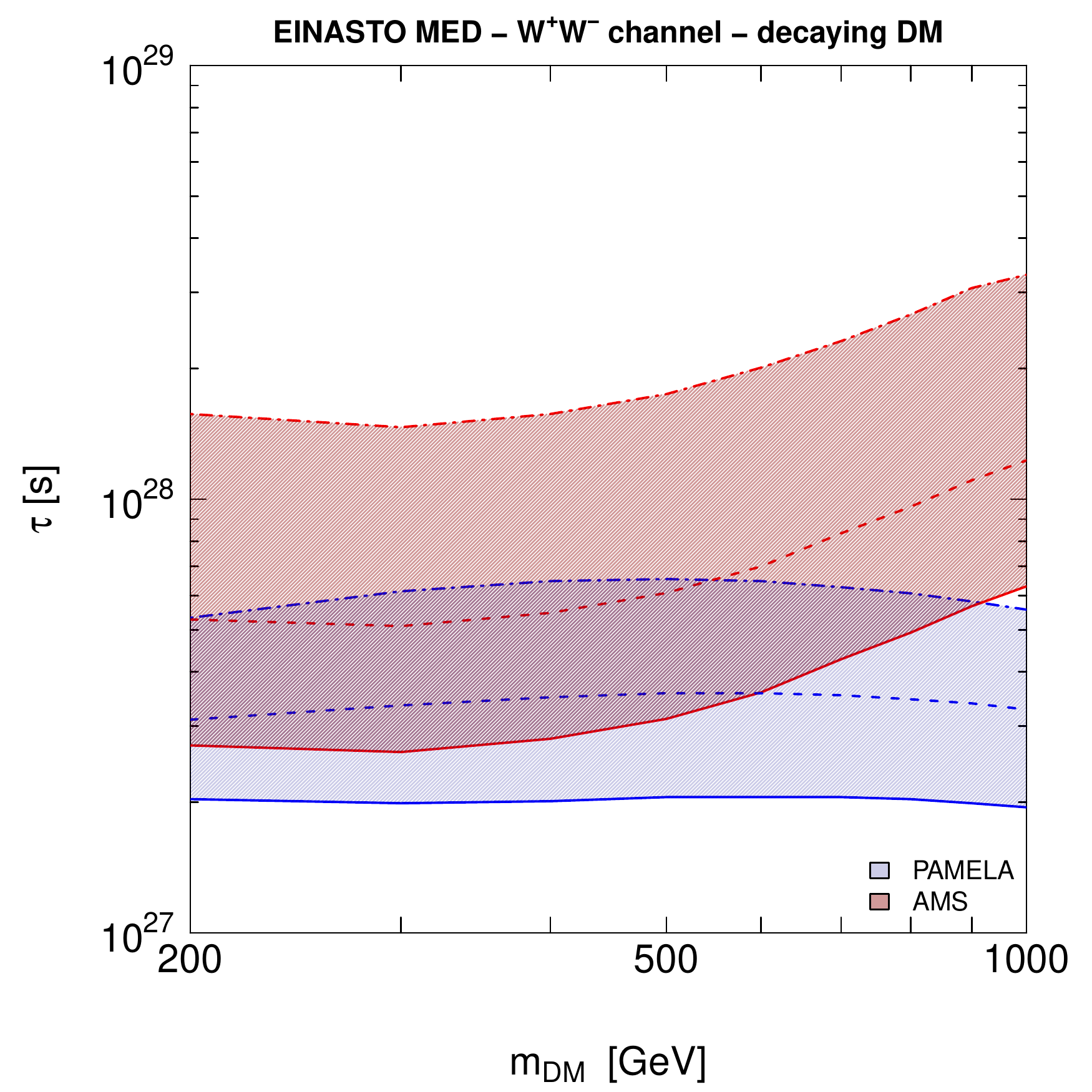}
\caption{The same as in Fig. \ref{fig:AMS_uu}, for the $W^+ W^-$ annihilation/decay channel.}
\label{fig:AMS_WW}
\end{figure}

Results are shown in Fig. \ref{fig:AMS_uu} for the $\bar u u$ production channel, 
in Fig. \ref{fig:AMS_bb} for the $\bar b b$ channel, and in Fig. \ref{fig:AMS_WW} for the 
$W^+W^-$ channel. The plots  show the projected sensitivity for AMS-02, for annihilating (left panel) and decaying (right panel) DM,
compared to the current bounds from PAMELA. The representative case reported in  
Fig. \ref{fig:AMS_uu}, \ref{fig:AMS_bb} and \ref{fig:AMS_WW} refers to
an Einasto density profile and the MED set
of  propagation parameters in the Galaxy. 
Each set of curves (in the left panel the ``upper'' blue band refers to PAMELA, the ``lower'' red band refers to AMS-02; the reverse occurs in the right panel: the `'lower'' blue band refers to PAMELA, the ``upper'' red band refers to AMS-02) show the current PAMELA bound or the projected AMS-02
sensitivity, under three different assumptions on the size of the theoretical uncertainties on the secondary
antiproton production: solid, dashed and dot-dashed lines refer to 40\%, 20\% and 5\%, respectively. The solid lines for PAMELA reproduce the bounds
 reported in Fig. \ref{fig:Pamela_MED}. The horizontal (green)
line in the left panel denotes the ``thermal'' value $\sigmav = 3\times 10^{-26}$ cm$^3$ s$^{-1}$.

First of all, we notice that the theoretical uncertainty on the background flux can represent
a dominant and limiting factor in the ability to improve the bounds on DM. By comparing the
current PAMELA limits and the AMS projected sensitivity obtained with a 40\% uncertainty
on the background flux (solid lines in Fig. \ref{fig:AMS_uu}, \ref{fig:AMS_bb} and \ref{fig:AMS_WW}) we see that AMS-02 will improve the bounds in the whole mass range and
for all antiproton production channels, but for DM masses below 100 GeV the improvement
will likely not be large. Only for DM masses above 100 GeV the bounds can be significantly
improved, mostly due to the fact that AMS-02 will have access to antiproton energies larger than
those covered by PAMELA. For very light DM, which produces antiprotons at low kinetic energies,
the geomagnetic cutoff can instead be a limiting factor: Fig. \ref{fig:AMS_uu} shows that for
DM lighter than a few GeV (which is a case relevant only for annihilation/decay into light quarks)
AMS-02 sensitivity drops. 

In the case theoretical uncertainties in the background flux can be reduced, both PAMELA bounds 
and AMS-02 projected sensitivities would improve. In this case, the larger statistics of AMS-02 could be more throughly exploited, and the expected reach significantly extended. This is manifest in Figs. \ref{fig:AMS_uu}, \ref{fig:AMS_bb} and \ref{fig:AMS_WW},
especially for a reduction of the theoretical uncertainties where both a 20\% level and a more
ambitious level of 5\% are reported, in which case an improvement of up to an order of magnitude
can be obtained, depending on the antiproton production channel and DM mass range.

\section{Conclusions}
 \label{sec:Conclusions} 

In this paper we have presented the most updated analysis of the bounds on DM properties that can be obtained from antiprotons measurements.  We have included in our analysis not only the uncertainties arising from galactic modeling (i.e. the DM density profile and, most relevant, the propagation parameters) which, as known, provide the largest variability in the derived bounds on DM properties,
but we have also investigated the impact of solar 
modulation modeling, 
 which we have shown to introduce an uncertainty typically of the order of 10-30\%, with
a maximal effect of about 50\%, with the largest impact occurring in the low DM mass range. To evaluate the importance of solar modulation, we have used a full numerical and charge-dependent solution of the equation that models cosmic rays transport
in the heliosphere, tuned on data sensitive to solar activity \cite{Maccione:2012cu}. This detailed modeling has allowed us to
quantify the impact of solar modulation on the derived bounds, once a galactic propagation model
is adopted.

We have shown that the constraining power of the antiprotons measurements for DM particles that annihilate into quarks or gauge bosons can be relevant: bounds on the DM annihilation cross section
(or lifetime, in the case of decaying DM) are strong, similar or in some cases even stronger than those that arise from gamma-ray measurements. Considering the most probable set of galactic propagation parameters (the MED model), for annihilating DM and "thermal" cross section the
whole DM mass range below 90 GeV is excluded, when DM annihilates into light quarks; this
bounds moves to 40 GeV when annihilation occurs into heavy quarks. In the case of decaying DM,
 the lower limit on the lifetime is set to $10^{28}$ s for intermediate DM masses and can reach  $10^{29}$ s for very light DM particles annihilating into light quarks. 
Concerning solar modulation, variations of the modeling parameters, in particular the value of the mean free path $\lambda$, have an impact on the bounds that can be as large as 30-50\% for the lightest DM particles and decreases as the DM particle mass grows. 
While these variations due to solar modulation modeling are not as large as those due to galactic
transport modeling, nevertheless they have a size that can influence the ability to set bounds
on the mass of annihilating DM: the quoted limit of 40 GeV for the mass of a DM particle
 annihilating into heavy quarks
can be varied in a range of values which extends up to 60 GeV, when solar modulation modeling is taken into account.

In the last section of the paper, we have investigated the future perspectives for antiproton searches
in the light of the AMS mission. We have shown that (and quantified how much) a high-precision experiment like AMS-02 will allow to set stronger bounds on DM properties, even if effects such as the geomagnetic cutoff can play a non-negligible role, since they can limit the sensitivity in the lower DM masses region. However, in order to fully exploit the AMS increased sensitivity, a reduction of the theoretical errors (mostly related to nuclear uncertainties in 
the antiproton production processes and to the determination  of the primary cosmic rays fluxes) in the calculation
of the astrophysical secondary antiproton background will be critically important.

\acknowledgments

N.F. and A.V. acknowledge the Research Grant funded by the Istituto Nazionale di Fisica Nucleare within the {\sl Astroparticle Physics Project}
(INFN grant code: FA51).  N.F. and A.V. acknowledge support of the {\sl Strategic Research Grant} jointly funded by the University of Torino and Compagnia di San Paolo 
(grant Unito-CSP 2010-2012 entitled {\sl Origin and Detection of Galactic and Extragalactic Cosmic Rays}). N.F. acknowledges support of the
spanish MICINN Consolider Ingenio 2010 Programme under grant MULTIDARK
CSD2009--00064. 
LM acknowledges support from the Alexander von Humboldt foundation and partial support from the European Union FP7 ITN INVISIBLES (Marie Curie Actions, PITN-GA-2011-289442).
This work is supported in part by the European Research Council ({\sc ERC}) under
the EU Seventh Framework Programme (FP7/2007-2013)/{\sc ERC} Starting Grant
(agreement n.\ 278234 --- `{\sc NewDark}' project).




  


\begin{thebibliography}{99}

\bibitem{Ade:2013zuv}
  P.~A.~R.~Ade {\it et al.}  [Planck Collaboration],
  \emph{Planck 2013 results. XVI. Cosmological parameters},
  \href{http://arxiv.org/abs/arXiv:1303.5076}{[arXiv:1303.5076 [astro-ph.CO]]}
      
\bibitem{Donato:1999gy}
  F.~Donato, N.~Fornengo and P.~Salati,
  \emph{Anti-deuterons as a signature of supersymmetric dark matter},
  Phys.\ Rev.\ D {\bf 62} (2000) 043003
  \href{http://arxiv.org/abs/hep-ph/9904481}{[arXiv:hep-ph/9904481]} 
      
\bibitem{Orito:1999re}
S.Orito et al. [BESS Collaboration]
\emph{Precision measurement of cosmic ray anti-proton
                        spectrum},
Phys.Rev.Lett. {\bf 84} (2000) 1078-1081
\href{http://arxiv.org/abs/astro-ph/9906426}{[arXiv:astro-ph/9906426]} 

\bibitem{Maeno:2000qx}
T.Maeno et al. [BESS Collaboration]
\emph{Successive measurements of cosmic ray anti-proton
                        spectrum in a positive phase of the solar cycle},
Astrop.Phys. {\bf 16} (2001) 121-128  
\href{http://arxiv.org/abs/astro-ph/0010381}{[arXiv:astro-ph/0010381]}     

\bibitem{Aguilar:2002ad}  
M. Aguilar et al. [AMS Collaboration]
\emph{The Alpha Magnetic Spectrometer (AMS) on the
                        International Space Station. I: Results from the test
                        flight on the space shuttle},
Phys.Rept. {\bf 366} (2002) 331-405 

\bibitem{Abe:2011nx}
K.Abe, H.Fuke, S.Haino, T.Hams, M. Hasegawa et al. 
\emph{Measurement of the cosmic-ray antiproton spectrum at
                        solar minimum with a long-duration balloon flight over
                        Antarctica}, 
Phys.Rev.Lett. {\bf 108} (2012) 051102
\href{http://arxiv.org/abs/arXiv:1107.6000}{[arXiv:1107.6000 [astro-ph]]}   

\bibitem{Adriani:2010rc}
  O.~Adriani {\it et al.}  [PAMELA Collaboration],
  \emph{PAMELA results on the cosmic-ray antiproton flux from 60 MeV to 180 GeV in kinetic energy},
  Phys.\ Rev.\ Lett.\  {\bf 105} (2010) 121101
  [arXiv:1007.0821 [astro-ph.HE]] 
  
\bibitem{Adriani:2012paa}
O. Adriani, G.A. Bazilevskaya, G.C. Barbarino, R. Bellotti, M.Boezio et al. 
\emph{Measurement of the flux of primary cosmic ray
                        antiprotons with energies of 60-MeV to 350-GeV in the
                        PAMELA experiment},
 JETP Lett. {\bf 96} (2013) 621-627   

\bibitem{Silk:1984zy}
J. Silk and M.Srednicki
\emph{Cosmic Ray anti-Protons as a Probe of a Photino
                        Dominated Universe},
                        Phys. Rev. Lett. {\bf 53} (1984) 624
                        
\bibitem{Stecker:1985jc}
F.W. Stecker, S. Rudaz and T.F. Walsh, 
\emph{Galactic Anti-protons From Photinos},
 Phys.Rev.Lett {\bf 55} (1985) 2622-2625  
 
 \bibitem{Jungman:1993yn}
 G. Jungman and M. Kamionkowski,
 \emph{Cosmic ray anti-protons from neutralino annihilation
                        into gluons},
                        Phys. Rev. {\bf D49} (1994) 2316-2321
                          \href{http://arxiv.org/abs/astro-ph/9310032}{[arXiv:astro-ph/9310032]}  
                        
 
 \bibitem{Bottino:1994xs}
 A.Bottino, C.Favero, N. Fornengo and G. Mignola,
 \emph{Amount of anti-protons in cosmic rays due to halo
                        neutralino annihilation}, 
 Astropart. Phys. {\bf 3} (1995) 77-86
  \href{http://arxiv.org/abs/hep-ph/9408392}{[arXiv:hep-ph/9408392]}   
  
  \bibitem{Bottino:1998tw}
  A.Bottino, F.Donato, N.Fornengo and P.Salati,
  \emph{Which fraction of the measured cosmic ray anti-protons
                        might be due to neutralino annihilation in the galactic
                        halo?},
                        Phys. Rev. {\bf D58} (1998) 123503 
   \href{http://arxiv.org/abs/astro-ph/9804137}{[arXiv:astro-ph/9804137]}                                         
 
 \bibitem{Bergstrom:1999jc}
 L. Bergstrom, J. Edsjo and P.Ullio ,
 \emph{Cosmic anti-protons as a probe for supersymmetric dark
                        matter?},
                        Atrophys. J. {\bf 526} (1999) 215-235
  \href{http://arxiv.org/abs/astro-ph/9902012}{[arXiv:astro-ph/9902012]}   
  
  \bibitem{Belanger:2012ta}
  G. Belanger, C. Boehm, M. Cirelli, J. Da Silva and A. Pukhov, 
  \emph{PAMELA and FERMI-LAT limits on the neutralino-chargino
                        mass degeneracy},           
   JCAP {\bf 1211} (2012) 028
   \href{http://arxiv.org/abs/1208.5009}{[arXiv:1208.5009 [hep-ph]]}                                                 

\bibitem{Donato:2003xg}
F. Donato, N. Fornengo,  D.Maurin and P. Salati
\emph{Antiprotons in cosmic rays from neutralino
                        annihilation},
                        Phys.Rev. {\bf D69} (2004) 063501
                        \href{http://arxiv.org/abs/astro-ph/0306207}{[arXiv:astro-ph/0306207]}
                    
\bibitem{Bottino:2005xy}
A. Bottino, F. Donato, N. Fornengo and P. Salati
\emph{Antiproton fluxes from light neutralinos},
Phys. Rev. {\bf D72} (2005) 083518
\href{http://arxiv.org/abs/hep-ph/0507086}{[arXiv:hep-ph/0507086]}    

\bibitem{Ferrer:2006hy}
F. Ferrer, L. Krauss and S.Profumo
\emph{Indirect detection of light neutralino dark matter in
                        the NMSSM},
                        Phys. Rev. {\bf D74} (2006) 115007
\href{http://arxiv.org/abs/hep-ph/0609257}{[arXiv:hep-ph/0609257]}   

\bibitem{Bottino:2007qg}
A. Bottino, F. Donato, N. Fornengo and S. Scopel
\emph{Zooming in on light relic neutralinos by direct
                        detection and measurements of galactic antimatter},
                        Phys. Rev. {\bf D77} (2008) 015002 
\href{http://arxiv.org/abs/0710.0553}{[arXiv:0710.0553 [hep-ph]]}     

\bibitem{Cerdeno:2011tf}
D. Cerdeno T. Delahaye and J. Lavalle
\emph{Cosmic-ray antiproton constraints on light singlino-like
                        dark matter candidates},     
                        Nucl. Phys {\bf B854}  (2011) 738-779 
                        \href{http://arxiv.org/abs/1108.1128}{[arXiv:1108.1128 [hep-ph]]}   
                        
\bibitem{Ibarra:2008qg}
A. Ibarra and D.Tran 
\emph{Antimatter Signatures of Gravitino Dark Matter Decay},
JCAP {\bf 0807} (2008) 002
\href{http://arxiv.org/abs/0804.4596}{[arXiv:0804.4596 [hep-ph]]} 
                        
\bibitem{Buchmuller:2009xv}    
W. Buchmuller, A.Ibarra, T.Shindou, F. Takayama and D. Tran, 
\emph{Probing Gravitino Dark Matter with PAMELA and Fermi},
JCAP {\bf 0909} (2009) 021 
\href{http://arxiv.org/abs/0906.1187}{[arXiv:0906.1187 [hep-ph]]}                   

      \bibitem{Evoli:2011id}
      C. Evoli, I. Cholis, D. Grasso, L. Maccione and P. Ullio, 
      \emph{Antiprotons from dark matter annihilation in the Galaxy: theoretical uncertainties},
      Phys. Rev. D {\bf 85} (2012) 123511
      \href{http://arxiv.org/abs/arXiv:1108.0664}{[arXiv:1108.0664 [astro-ph]]}         
                        
\bibitem{Delahaye:2013yqa}
T. Delahaye and M.Grefe,
\emph{Antiproton Limits on Decaying Gravitino Dark Matter},
2013 \href{http://arxiv.org/abs/1305.7183}{[arXiv:1305.7183]}                        
                        
                        
\bibitem{Bringmann:2005pp}
T. Bringmann 
\emph{High-energetic cosmic antiprotons from Kaluza-Klein dark
                        matter},
                        JCAP {\bf 0508} (2005) 006
                        \href{http://arxiv.org/abs/astro-ph/0506219}{[arXiv:astro-ph/0506219]}     
                        
\bibitem{Barrau:2005au}
A.Barrau, P.Salati, G.Servant, F.Donato, J.Grain et al. 
\emph{Kaluza-Klein dark matter and Galactic antiprotons},
Phys. Rev. {\bf D72} (2005) 063507
\href{http://arxiv.org/abs/astro-ph/0506389}{[arXiv:astro-ph/0506389]}   

\bibitem{Hooper:2009fj}
D. Hooper, K.M. Zurek, 
\emph{The PAMELA and ATIC Signals From Kaluza-Klein Dark
                        Matter},    
Phys. Rev. {\bf D79} (2009) 103529
\href{http://arxiv.org/abs/0902.0593}{[arXiv:0902.0593 [hep-ph]]}  

\bibitem{Cirelli:2008id}
M. Cirelli, R. Franceschini and A. Strumia, 
\emph{Minimal Dark Matter predictions for galactic positrons,
                        anti-protons, photons},
Nucl. Phys{\bf B800} (2008) 204-220
\href{http://arxiv.org/abs/0802.3378}{[arXiv:0802.3378 [hep-ph]]}       

\bibitem{Garny:2011cj}
M. Garny, A. Ibarra and S. Vogl,
\emph{Antiproton constraints on dark matter annihilations from
                        internal electroweak bremsstrahlung},
JCAP {\bf 1107} (2011) 028
 \href{http://arxiv.org/abs/1105.5367}{[arXiv:1105.5367 [hep-ph]]} 
 
 \bibitem{Chu:2012qy}
 X. Chu, T. Hambye, T. Scarna and M.H.G. Tytgat
 \emph{What if Dark Matter Gamma-Ray Lines come with Gluon
                        Lines?},
                        Phys. Rev. {\bf D86} (2012) 083521
                        \href{http://arxiv.org/abs/1206.2279}{[arXiv:1206.2279 [hep-ph]]} 
 
\bibitem{Ibarra:2012dw}
A. Ibarra, S. Lopez Gehler  and M. Pato,
\emph{Dark matter constraints from box-shaped gamma-ray
                        features},
 JCAP {\bf 1207} (2012) 043 
 \href{http://arxiv.org/abs/1205.0007}{[arXiv:1205.0007 [hep-ph]]}                      

\bibitem{Lavalle:2010yw}
J. Lavalle, 
\emph{10 GeV dark matter candidates and cosmic-ray
                        antiprotons},
                        Phys. Rev. {\bf D82}(2010)
                        \href{http://arxiv.org/abs/1108.1128}{[arXiv:1007.5253 [astro-ph]]}  
                        
                         \bibitem{Donato:2008jk}
  F. Donato, D. Maurin, P. Brun, T. Delahaye and P. Salati, 
  \emph{Constraints on WIMP Dark Matter from the High Energy PAMELA $\bar{p}/p$ data},
  Phys. Rev. Lett. {\bf 102} (2009) 071301
  \href{http://arxiv.org/abs/0810.5292}{[arXiv:0810.5292 [astro-ph]]} 
  
  \bibitem{Kappl:2011jw}
  R. Kappl and M.W. Winkler 
  \emph{Dark Matter after BESS-Polar II},
  Phys. Rev. {\bf D85} (2012) 123522
  \href{http://arxiv.org/abs/1110.4376}{[arXiv:1110.4376 [hep-ph]]} 
  
  \bibitem{Garny:2012vt}
  M. Garny, A. Ibarra and D. Tran,
  \emph{Constraints on Hadronically Decaying Dark Matter},
  JCAP {\bf 1208} (2012) 025
  \href{http://arxiv.org/abs/1205.6783}{arXiv:1205.6783 [hep-ph]}
  
    \bibitem{Cirelli:2013hv}
  M.Cirelli and G. Giesen, 
  \emph{Antiprotons from Dark Matter: Current constraints and
                        future sensitivities}, 
      JCAP {\bf 1304} (2013) 015
      \href{http://arxiv.org/abs/arXiv:1301.7079}{arXiv:1301.7079 [hep-ph]}
                                                                                                                                                 

\bibitem{Fornengo:2013osa}
  N. Fornengo, L. Maccione, A. Vittino, 
  \emph{Dark matter searches with cosmic antideuterons: status and perspectives},
  JCAP {\bf 09} (2013) 031
  \href{http://arxiv.org/abs/arXiv:1306.4171}{[arXiv:1306.4171 [hep-ph]]}
  
  \bibitem{Sjostrand:2007gs}
  T. Sjostrand, S. Mrenna and P.Z. Skands
  \emph{A Brief Introduction to PYTHIA 8.1},
  Comput. Phys. Commun. {\bf 178} (2008) 852-867
  \href{http://arxiv.org/abs/arXiv:0710.3820}{[arXiv:0710.3820 [hep-ph] ]}
  
  \bibitem{Maurin:2002ua}
Maurin, David and Taillet, Richard and Donato, Fiorenza
                        and Salati, Pierre and Barrau, Aurelien and others,
                        \emph{Galactic cosmic ray nuclei as a tool for astroparticle
                        physics}
                        \href{http://arxiv.org/abs/astro-ph/0212111}{[arXiv:astro-ph/0212111]}
  
  \bibitem{diffusion1}
D. Maurin, F. Donato, R. Taillet and P. Salati, 
\emph{Cosmic rays below z=30 in a diffusion model: new constraints on propagation parameters},
Astrophys. J., {\bf 555} (2001) 585-596 
\href{http://arxiv.org/abs/astro-ph/0101231}{[arXiv:astro-ph/0101231]}

         \bibitem{diffusion2}
F. Donato, D. Maurin and R. Taillet,
\emph{Beta-radioactive cosmic rays in a diffusion model: test for a local bubble?},
Astron. Astrophys. {\bf 381} (2002) 539-559 
\href{http://arxiv.org/abs/astro-ph/0108079}{[arXiv:astro-ph/0108079]}

       \bibitem{diffusion3}
D. Maurin, R. Taillet and F. Donato,
\emph{New results on source and diffusion spectral features of galactic cosmic rays: I- B/C ratio},
Astron. Astrophys. {\bf 394} (2002) 1039-1056 
\href{http://arxiv.org/abs/astro-ph/0206286}{[arXiv:astro-ph/0206286]}


                        
                          
  \bibitem{Donato:2001ms}
  F.~Donato, D.~Maurin, P.~Salati, A.~Barrau, G.~Boudoul and R.~Taillet,
  Astrophys.\ J.\  {\bf 563} (2001) 172
  \href{http://arxiv.org/abs/astro-ph/0103150}{[arXiv: astro-ph/0103150]}
                                       
 \bibitem{Gleeson_1968ApJ} Gleeson, L.~J., \& Axford, W.~I.,\ 
\emph{Solar Modulation of Galactic Cosmic Rays},
Ap. J., {\bf 154}, (1968) 1011 

\bibitem{1996ApJ...464..507C} Clem, J.~M., Clements, D.~P., Esposito, J., et al.,\ 
\emph{Solar Modulation of Cosmic Electrons},
 Ap. J. {\bf 464} (1996) 507 
 
\bibitem{wilcox} \url{http://wso.stanford.edu/}

\bibitem{1981JGR....86.8893B} Burlaga, L.~F., Hundhausen, A.~J., \& Zhao, X.-P.,\ 
\emph{The coronal and interplanetary current sheet in early 1976},
Journal of Geophysical Research {\bf 86} (1981) 8893
 
\bibitem{1965P&SS...13....9P} Parker, E.~N.,\ 
\emph{The passage of energetic charged particles through interplanetary space},
P\&SS  {\bf 13} (1965) 9 

\bibitem{Bobik:2011ig} 
  P.~Bobik, G.~Boella, M.~J.~Boschini, C.~Consolandi, S.~Della Torre, M.~Gervasi, D.~Grandi and K.~Kudela {\it et al.},
  \emph{Systematic Investigation of Solar Modulation of Galactic Protons for Solar Cycle 23 using a Monte Carlo Approach with Particle Drift Effects and Latitudinal Dependence},
  Astrophys.\ J.\  {\bf 745} (2012) 132 
  \href{http://arxiv.org/abs/1110.4315}{[arxiv:1110.4315 [astro-ph.SR]]}



\bibitem{2011ApJ...735...83S} Strauss, R.~D., Potgieter, M.~S., B{\"u}sching, I., \& Kopp, A.,\ 
\emph{Modeling the Modulation of Galactic and Jovian Electrons by Stochastic Processes}, Ap. J. {\bf 735} (2011) 83 

\bibitem{2012Ap&SS.339..223S} Strauss, R.~D., Potgieter, M.~S., B{\"u}sching, I., \& Kopp, A.,\ 
\emph{Modelling heliospheric current sheet drift in stochastic cosmic ray transport models}, Ap\&SS {\bf 339} (2012)  223 


\bibitem{1999ApJ...520..204G} Giacalone, J., \& Jokipii, J.~R.,\ 
\emph{The Transport of Cosmic Rays across a Turbulent Magnetic Field},
 Ap. J.  {\bf 520}  (1999) 204 
 
\bibitem{1977ApJ...213L..85J} Jokipii, J.~R., \& Levy, E.~H.,\ 
\emph{Effects of particle drifts on the solar modulation of galactic cosmic rays},
Ap. J. Lett.  {\bf 213}  (1977) L85 

\bibitem{GastSchael} Gast, H. and Schael, S., Proc. of the 31st ICRC, Lodz, 2009
\bibitem{DellaTorre:2012zz} S. Della Torre {\em et al.}, 2012, Advances in Space Research, 49, 1587 
\bibitem{Potgieter:2013cwj} 
  M.~S.~Potgieter, E.~E.~Vos, M.~Boezio, N.~De Simone, V.~Di Felice and V.~Formato,
  \emph{Modulation of galactic protons in the heliosphere during the unusual solar minimum of 2006 to 2009},
  \href{http://arxiv.org/abs/1302.1284}{[arxiv:1302.1284 [astro-ph.SR]]}

 

\bibitem{Maccione:2012cu} 
L.~Maccione,
  \emph{Low energy cosmic ray positron fraction explained by charge-sign dependent solar modulation},
  Phys.\ Rev.\ Lett.\  {\bf 110}, 081101 (2013)
  \href{http://arxiv.org/abs/1211.6905}{[arxiv:1211.6905 [astro-ph.HE]]}
  
\bibitem{2007JGRA..11208101A}
K. Alanko-Huotari, I. G. Usoskin, K. Mursula, and G. A. Kovaltsov, 
\emph{Stochastic simulation of cosmic ray modulation including a wavy heliospheric current sheet},
Journal of Geophysical Research (Space Physics) {\bf 112} (2007) A08101.

\bibitem{gardiner2009stochastic}
C. Gardiner, 
\emph{Stochastic Methods:A Handbook for the Natural and Social Sciences}, 
Springer Series in Synerget-ics (Springer, 2009).

\bibitem{2012CoPhC.183..530K}
A. Kopp, I. B\"usching, R. D. Strauss, and M. S. Potgieter, 
\emph{A stochastic differential equation code for multidimensional FokkerÐPlanck type problems},
Computer Physics Communications {\bf 183} (2012) 530.

\bibitem{Potgieter2003}
S. Ferreira, M. Potgieter, 
\emph{Modulation over a 22-year cosmic ray cycle: On the tilt angles of the heliospheric current sheet},
Advances in Space Research {\bf 32} (2003) 657.

\bibitem{Potgieter2004}
S. Ferreira, M. S. Potgieter, 
\emph{Long-Term Cosmic-Ray Modulation in the Heliosphere}
ApJ 603 (2004) 744.

\bibitem{Droge2005532} Dr{\"o}ge, W.\  
\emph{Probing heliospheric diffusion coefficients with solar energetic particles},
Advances in Space Research {\bf 35} (2005) 532 

 \bibitem{Loparco:2013pea}
  F.~Loparco, L.~Maccione and M.~N.~Mazziotta,
  \emph{Inferring the local interstellar spectrum of cosmic ray protons from PAMELA data},
  \href{http://arxiv.org/abs/1306.1354}{[arXiv:1306.1354 [astro-ph.HE]]}
  
  \bibitem{gammabound1}
A. Abdo {\it et al.} [Fermi-LAT Collaboration],
\emph{Observations of Milky Way Dwarf Spheroidal galaxies with the Fermi-LAT
detector and constraints on Dark Matter models},
  Astrophys. J. {\bf 712} (2010) 147
  \href{http://arxiv.org/abs/1001.4531}{[arXiv:1001.4531 [astro-ph.CO]]}


\bibitem{gammabound2}
M. Ackermann {\it et al.} [Fermi-LAT Collaboration],
\emph{Constraints on Dark Matter Annihilation in
Clusters of Galaxies with the Fermi Large Area
Telescope},
  JCAP {\bf 5} (2010) 025
  \href{http://arxiv.org/abs/1002.2239}{[arXiv:1002.2239 [astro-ph.CO]]}

\bibitem{gammabound3}
A. Abdo {\it et al.} [Fermi-LAT Collaboration],
\emph{Constraints on Cosmological Dark Matter Annihilation from the Fermi-LAT
Isotropic Diffuse Gamma-Ray Measurement},
  JCAP {\bf 4} (2010) 014
  \href{http://arxiv.org/abs/1002.4415}{[arXiv:1002.4415 [astro-ph.CO]]}

\bibitem{gammabound4}
A. Geringer-Sameth, S.M. Koushiappas,
\emph{Exclusion of canonical WIMPs by the joint analysis of Milky Way dwarfs with
Fermi},
  Phys.\ Rev.\ Lett.\  {\bf 107} (2011) 241303
  \href{http://arxiv.org/abs/1108.2914}{[arXiv:1108.2914 [astro-ph.CO]]}

\bibitem{gammabound5}
M. Ackermann {\it et al.} [Fermi-LAT Collaboration],
\emph{Constraining Dark Matter Models from a Combined Analysis of Milky Way
Satellites with the Fermi Large Area Telescope},
  Phys.\ Rev.\ Lett.\  {\bf 107} (2011) 241302
  \href{http://arxiv.org/abs/1108.3546}{[arXiv:1108.3546 [astro-ph.HE]]}

\bibitem{gammabound6}
E. Nezri, R. White, C. Combet, J.A. Hinton, D. Maurin, E. Pointecouteau,
\emph{$\gamma$-rays from annihilating dark matter in galaxy clusters: stacking vs
single source analysis},
  MNRAS {\bf 425} (2012) 477
  \href{http://arxiv.org/abs/1203.1165}{[arXiv:1203.1165 [astro-ph.HE]]}

\bibitem{gammabound7}
M. Ackermann {\it et al.} [Fermi-LAT Collaboration],
\emph{Constraints on the galactic halo dark matter from fermi-lat diffuse
measurements},
  Astrophys. J. {\bf 761} (2012) 91
  \href{http://arxiv.org/abs/1205.6474}{[arXiv:1205.6474 [astro-ph.CO]]}


\bibitem{gammabound8}
T. Bringmann, F. Calore, M. Di Mauro, F. Donato, 
\emph{Constraining dark matter annihilation with the isotropic $\gamma$-ray
background: updated limits and future potential},
Phys. Rev. {\bf D 89} (2014) 023012
  \href{http://arxiv.org/abs/1303.3284}{[arXiv:1303.3284 [astro-ph.CO]]}
  
    
   \bibitem{AMS_acc}
   A. G. Malinin [AMS Collaboration], 
   \emph{Astroparticle physics with AMS-02},
   Phys. Atom. Nucl. {\bf 67} (2004) 2044



\bibitem{geo_battiston}
R. Battiston, 
\emph{Recent results from the AMS experiment on the International Space Station},
XVI Lomonosov Conference on Elementary Particle Physics, Moscow State
University, Moscow (Russia), 22-28 August 2013.

  \end{thebibliography}
\end{document}